\documentclass[%
reprint,
 amsmath,amssymb,
 aps,
]{revtex4-2}

\usepackage{graphicx}
\usepackage{dcolumn}
\usepackage{bm}
\usepackage[caption=false]{subfig}

\begin{document}

\preprint{APS/123-QED}

\title{Extreme heterogeneity in the microrheology of lamellar surfactant \\gels analyzed with neural networks}

\author{Owen Watts Moore}
\altaffiliation{Photon Science Institute, University of Manchester, Manchester M13 9PL, UK}
\author{Conor Lewis}
\altaffiliation{Photon Science Institute, University of Manchester, Manchester M13 9PL, UK}
\author{Thomas Ross}
\altaffiliation{Photon Science Institute, University of Manchester, Manchester M13 9PL, UK}
\author{Thomas Andrew Waigh}
\altaffiliation{Photon Science Institute, University of Manchester, Manchester M13 9PL, UK}
\email{t.a.waigh@manchester.ac.uk}

\affiliation{Biological Physics, Department of Physics and Astronomy, University of Manchester, Manchester M13 9PL,
UK
}%
\author{Nickolay Korabel}
\affiliation{Department of Mathematics, The University of Manchester, Manchester M13 9PL, UK
}%
\author{Cesar Mendoza}
\affiliation{Unilever Research \(\&\) Development, Port Sunlight Laboratory, Quarry Road East, Bebington, Wirral CH63 3JW, UK
}%

\begin{abstract}
The heterogeneity of the viscoelasticity of a lamellar gel network based on cetyl-trimethylammonium chloride (CTAC) and ceto-stearyl alcohol was studied using particle tracking microrheology. A recurrent neural network (RNN) architecture was used for estimating the Hurst exponent, \(H\), on small sections of tracks of probe spheres moving with fractional Brownian motion. Thus dynamic segmentation of tracks via neural networks was used in microrheology for the first time and it is significantly more accurate than using mean square displacements. An ensemble of 414 particles produces a mean squared displacement (MSD) that is subdiffusive in time, \(t\), with a power law of the form \(t^{0.74\pm0.02}\), indicating power law viscoelasticity. RNN analysis of the probability distributions of \(H\), combined with detailed analysis of the time-averaged MSDs of individual tracks, revealed diverse diffusion processes belied by the simple scaling of the ensemble MSD, such as caging phenomena, which give rise to the complex viscoelasticity of lamellar gels.
\end{abstract}

\maketitle

\section{\label{sec:introduction}Introduction}
Particle tracking microrheology (PTM) has emerged in the past 30-40 years as a standard tool for probing the microstructure and linear rheology of soft viscoelastic fluids  \cite{Waigh2005}. Compared to bulk rheology experiments, a much larger frequency range can be accessed in a single PTM experiment, while using much less fluid \cite{Mason1995}. PTM techniques are diverse in their methodology, but all use the random, thermally driven, motion of probe particles suspended in a solvent to discern the rheological properties of the sample \cite{Waigh2005}. Modern PTM techniques can broadly be split into two categories: passive and active. In passive microrheology, the motion of the probe particles is studied in the absence of an external driving force. Such methods include diffusing wave spectroscopy (DWS) \cite{Mason1995}, laser deflection particle tracking (LDTP) \cite{Mason1997b}, and video PTM \cite{Mason1997a,Apgar2000,Tseng2001}. Active microrheology methods, such as magnetic tweezers \cite{Ziemann1994}, use an external force to move the probe particles, while the fluid-dependent response is recorded.
The fundamental measurement made in PTM experiments is the mean squared displacement (MSD), \(\langle r^2\rangle\), of the particles' motion \cite{Waigh2005}. In a viscous fluid, the MSD is related to time, \(t\) via the equation:
\begin{equation}\label{eq:SE eqn}
    \langle r^2(t)\rangle=2nDt,
\end{equation}
where \(D\) is the diffusion coefficient, and \(n\) is the number of dimensions \cite{Waigh2005,Bouchaud1990}. The use of the MSD in this context and its linearity in time was devised by Einstein in his landmark 1905 paper \cite{Einstein1905}. Ninety years later, Mason and Weitz (1995) derived a method for calculating the viscoelastic spectrum, \(\Tilde{G}(s)\), from \(\langle r^2\rangle\) in the Laplace domain, \cite{Mason1995} \begin{equation}\label{eq:shear mod}
    \Tilde{G}(s) = \frac{k_BT}{\pi a s \langle \Tilde{r}^2(s) \rangle},
\end{equation}
where \(s\) is the Laplace frequency, \(a\) is the tracer particle radius, \(k_BT\) is the thermal energy, and a tilde corresponds to a variable in the Laplace domain. This an expression of the generalised Stoke-Einstein equation and neglects an inertial term that is significant at high frequencies. One consequence of equation \ref{eq:shear mod} is that the creep compliance, \(J(t)\), is linearly related to \(\langle r^2\rangle\) by a constant of proportionality, without any need to change domain \cite{Petka1998,Mason2000},
\begin{equation}\label{eq:compliance}
    J(t) = \frac{\pi a^2\langle r^2(t)\rangle}{k_BT}.
\end{equation}
So long as the size of the probe particles are larger than the largest structure in the fluid \cite{Mason1995}, this directly links the MSD to bulk linear rheological properties of the fluid. This is often inaccessible in complex fluids for mechanical rheometers as fluid microstructures can be fragile and sensitive to deformation. As a result, PTM has been used in a huge variety of media including living cells \cite{Harrison2013,Caspi2000,Rogers2008,Wirtz2009}, other biological substances \cite{Mason1997a,Georgiades2014,Gardel2003,Tseng2001,Apgar2000,Ziemann1994,Hart2019,Hasnain2006}, and many non-biological complex fluids \cite{Mason1995,Lavrentovich2014,Ganapathy2007,Mason1997b}. For reviews of PTM, see references \cite{Waigh2005,Waigh2016}.\\

Fluids that do not obey equation \ref{eq:SE eqn} are said to show anomalous diffusion \cite{Bouchaud1990}. A common deviation is for the MSD to follow a power law of the form,
\begin{equation}\label{eq:anomalous diffusion}
    \langle r^2(t)\rangle \propto t^\alpha, \; \; 0<\alpha<2.
\end{equation} 
When \(\alpha = 1\), the linearity of equation \ref{eq:SE eqn} for viscous fluids is recovered. The \(\alpha<1\) case is labelled subdiffusion, while \(\alpha>1\) is superdiffusion \cite{Bouchaud1990}. Despite its name, anomalous diffusion is incredibly common throughout the sciences \cite{Klafter2005,Waigh2023}, and naturally arises in stochastic processes defined by the sum of many microscopic events, where the macro-statistics are invariant to the micro-statistics \cite{Eliazar2011}. In fluids, subdiffusion can arise from physical or energy barriers temporarily capturing tracer particles or the intrinsic viscoelasticity of the tracer-fluid mixture \cite{Sokolov2012}. Superdiffusion usually requires a driving force. As such, it can often be found in living systems \cite{Caspi2000,Harrison2013}. In inanimate non-driven systems, the average behaviour at long times is diffusion or subdiffusion to avoid breaking the second law of thermodynamics. For example, it has been detected when associated with fluctuating quantities in complex fluids such as wormlike micelles \cite{Ott1990,Ganapathy2007,Gambin2005}.\\

Two of the most common models for anomalous diffusion are the continuous time random walk (CTRW) and fractional Brownian motion (fBm) \cite{Montroll1965, Mandelbrot1968}. Though the time dependence of the MSD is similar (equation \ref{eq:anomalous diffusion}), their mechanisms and the real-world manifestation of their statistics are quite different \cite{Magdziarz2008,Burov2011}. A CTRW is based on a particle taking discrete jumps. In a decoupled case, it is governed by two independent probability distributions for the jump vector and the time before a jump is made, \(\tau\) \cite{Metzler2000}. If the mean of the \(\tau\) distribution is well-defined, the MSD will follow equation \ref{eq:SE eqn}. If the mean diverges, the process will occur at a slower rate and will be subdiffusive, leading to equation \ref{eq:anomalous diffusion} \cite{Sokolov2012,Sokolov2005,Klafter2005}. The distribution in the latter case is non-stationary as large values of \(\tau\) correspond to a lower probability of making a jump. fBm, $B_H(t)$, is a generalisation of Brownian motion to include correlated steps \cite{Eliazar2013}. The covariance function of fBm has the form $\mathbb{E}[B_H(t)B_H(s)]=\frac{1}{2}(|t|^{2H}+|s|^{2H}-|t-s|^{2H})$, such that for $t=s$, \(\langle r^2\rangle \propto t^{2H}\) with \(0<H<1\). Accordingly, the ranges \(0<H<1/2\) and \(1/2<H<1\) correspond to subdiffusion and superdiffusion respectively. In the language of fBm, these regimes are designated as \emph{anti-persistent}, where the steps are negatively correlated, and \emph{persistent}, where the steps are positively correlated. The only distribution governing fBm is the step vector, which is stationary for unobstructed motion with a single value of \(H\).\\

An MSD can also be produced via a time average, given by,
\begin{equation}\label{eq:TA MSD}
    \overline{r^2}(\Delta) = \frac{\int^{T-\Delta}_0[\textbf{r}(t+\Delta)-\textbf{r}(t)]^2dt}{T-\Delta}
\end{equation}
in the continuous case. The time lag, \(\Delta\), is the difference in time between two points in the random walk, while \(T\) is the total walk time. Throughout this paper, a bar above a variable indicates a time average (TA) and angled brackets an ensemble average (EA). TA MSDs provide better signal to noise than EA and are often preferred in situations where there are few tracks available. In Brownian fluids, if \(T\) is large, \(\overline{r^2}(\Delta)\) will tend to \(\langle r^2(t)\rangle\) \cite{Pusey1994,Lubelski2008}. This is an expression of ergodicity, where, given enough time, a random variable will sample the entire phase space of a system \cite{Rudnick2004}. Many fluids are non-ergodic. This can stem from locally varying rheological properties due to spatial heterogeneity \cite{Pusey1994,Jeon2010}, or an expression of non-stationary dynamics, such as that of \(\tau\) for the subdiffusive CTRW \cite{Lubelski2008,Bel2005,He2008,Burov2010,Jeon2010}. In heterogeneous fluids, probe particles will belong to sub-ensembles based on their environment. If particles are sampled proportionately from every sub-ensemble, then ergodicity will be recovered \cite{Pusey1994}. Since the only stochastic process governing fBm is stationary, fBm is ergodic \cite{Eliazar2013,Deng2009,Burov2011}. In complex fluids, fBm has been observed using single-particle tracking \cite{weiss2013,golding2006}.\\

\emph{Lamellar gel networks} (LGNs) are complex, multi-phase structures e.g. hair conditioners or pharmaceutical creams \cite{Iwata2017}. They contain a lamellar gel phase made up of bilayers of surfactant and fatty alcohol separated by water, bulk water, and crystals of hydrated fatty alcohol \cite{Junginger1984,Eccleston1997,Iwata2017}. At rest, their microstructure is highly heterogeneous and anisotropic, leading to elastic behaviour \cite{Hoffmann1996,Warriner1996,Iwata2017}. The rheology of such samples is very rich in phenomena and consequently poorly understood \cite{WattsMoore2023} e.g. hysteresis, shear banding and wall slip are observed as characteristic of soft-glassy rheological materials. This paper contains an analysis of the dynamics of an LGN based on cetyltrimethylammonium chloride (CTAC) using video particle tracking microrheology. Both time and ensemble averaged MSDs are used, revealing a diverse micro-environment producing subdiffusion on average. A recurrent neural network method for estimating the local Hurst exponent based on the feedforward neural network of Han et al. (2020) \cite{Han2020} is then used to analyse the local motion of the probe particles. We have combined single and multi output neuron models to improve the accuracy of the model for low Hurst exponents. The response of the combined model to simulations is first detailed before the method is applied to heterogeneous dynamics in LGN. Experiments have been chosen that highlight the rich phenomenology governing diffusion in lamellar gel networks.

\section{Materials and methods}
\subsection{Samples}
The LGN used in these experiments is based on a 3:1 ratio of CTAC to ceto-stearyl alcohol. Other components are: demineralised water, versene N\(_2\) crystals (a salt), and a preservative, Kathon CG. This was provided by Unilever and prepared according to the European patent by Casugbo et al. (2014) \cite{Casugbo2014Composition} and also Cunningham et al. (2021) \cite{Cunningham2021}. Identical samples were previously studied in optical coherence tomography non-linear rheology experiments \cite{WattsMoore2023}. To perform PTM, microparticles must be suspended in the fluid. For this purpose, carboxyl latex microbeads with a diameter of \(0.5 \;\mu\)m (ThermoFisher Scientific) and \(3\times10^8\) carboxyl groups per sphere surface were chosen. Particles larger than this were found to appear static due to the elasticity of the LGN, and the finite resolution of our microscope to measure particle displacement ($\sim$10 nm). If directly mixed with the LGN, the beads readily aggregate complicating the measurements. To avoid this, long chains of poly(ethylene glycol) (PEG) were added to the surface of the beads by covalent bonds with the surface carboxyl groups, a process called PEGylation. The exact method followed can be found on page 560 of Hermanson (2013) \cite{Hermanson2013a}. Hydrophilic PEG stabilises the particles in suspension by covering them in flexible polymers that provide a repulsive steric potential that is entropic in origin. This repulsive steric potential limits aggregation of particles \cite{Hermanson2013a,Jeon1991,Needham1992,Lasic1991}.

Once PEGylated, the spheres were added to the LGN. The elasticity of the LGN at rest precludes standard methods of homogenisation, such as sonication, or the use of a centrifuge. Instead, a vortex mixer was used for 5 minutes to fully suspend the spheres. 10-20 \(\mu\)l of the suspension was then placed on microscope coverslips and sealed with adhesive spacers. Once sealed, the sample was left for several hours before use to allow any residual mechanical stresses that may have been introduced during preparation to fully relax.

\subsection{Particle tracking Microrheology}
Samples were loaded onto an Olympus IX-71 inverted microscope and illuminated by a CoolLED pE-100 light source. The particles were viewed through a 100x immersion oil lens and their motion was recorded on a Photron Fastcam PCI camera. The whole microscope is mounted on an AVI350M dynamic vibration isolation system to reduce the impact of ambient vibrations combined with a sheet of acoustic isolation foam and a large floated optical table. Videos were taken at frame rates in the range 50-10,000 fps to access as large a range of time scales as possible. The particles were then tracked using an in-house program, \textit{PolyParticleTracker}, written in MATLAB by Rogers et al. (2007) \cite{Rogers2007}. This software fits a 2D, fourth-order polynomial, weighted by a Gaussian, to each particle. This tracking method is reasonably flexible with respect to particle shape and also intrinsically insensitive to the image background, while providing sub-pixel accuracy on particle position.

\subsection{Recurrent neural network}
Han et al. (2020) used a deep learning feedforward neural network (DLFNN) trained on fBm to study the intracellular motion of both endosomes and lysosomes \cite{Han2020} via dynamic segmentation of tracks. Building on this work, we have made a recurrent neural network (RNN) to estimate the Hurst exponent in 15-step segments of fBm tracks. RNNs are known to be well suited to sequential time series data due to their ability to remember past data which is crucial for non-Markovian processes \cite{Yu2019,Shewalkar2019}. Our RNN has 5 hidden layers, and a single output neuron that gives a continuous value for the estimated \(H\), \(H_{est}\). The model was trained on 10000 instances of 15-step fBm created with random simulated \(H\) exponents, \(H_{sim}\), generated using the Hosking method \cite{Hosking1984}. fBm was chosen over other anomalous diffusion processes because of its applicability on the micro-level and its compatibility with the regularly sampled time steps produced in video PTM experiments. The RNN model hyperparameters, including layer structure, activation function, number of neurons, and dropout rate were chosen via a Bayesian optimisation \cite{Wu2019} routine with 700 iterations, each using the same 10000 tracks. \\

\begin{figure}
\centering
    \includegraphics[width=0.9\columnwidth]{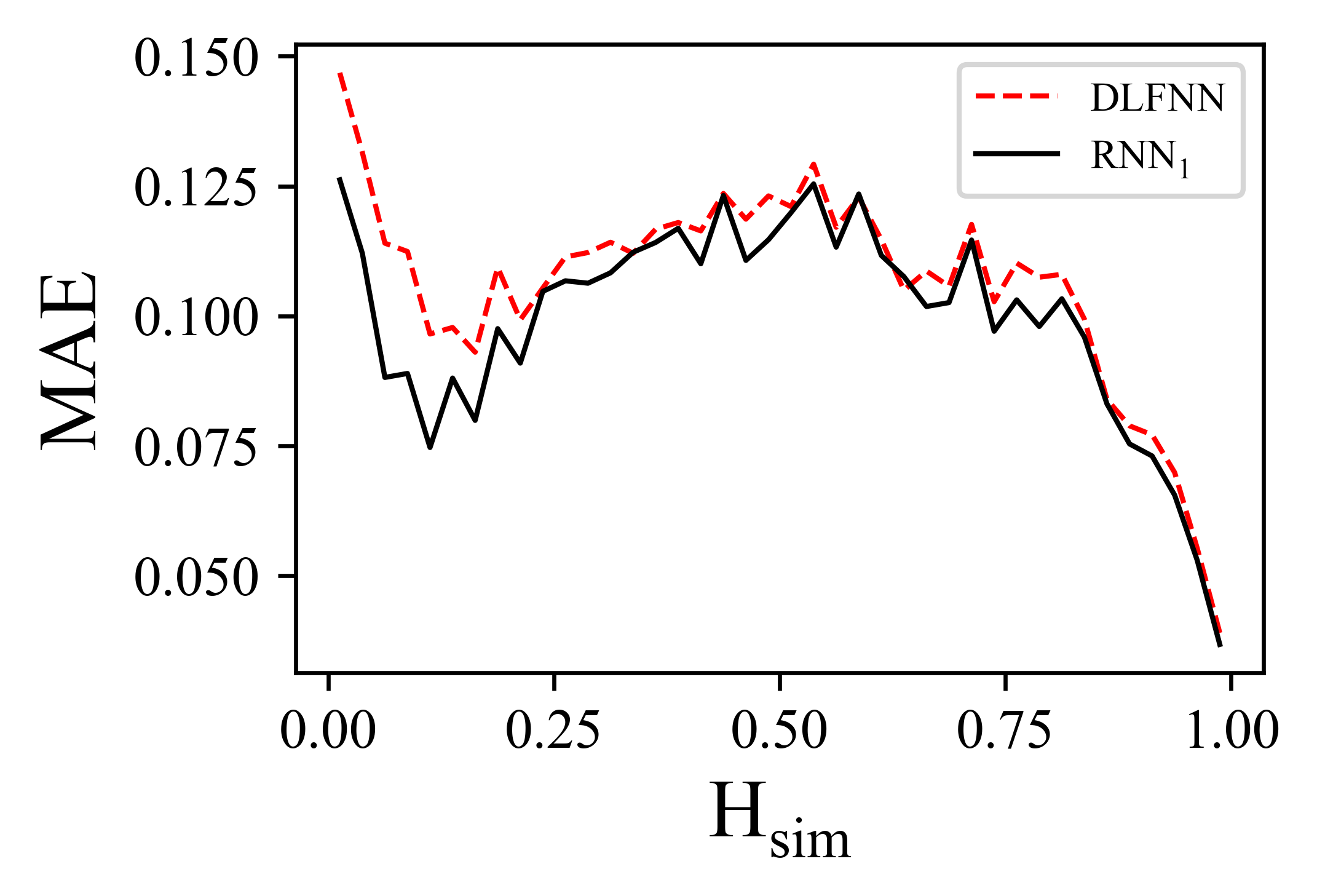}
    \caption{Mean absolute error (MAE) as a function of the simulated Hurst exponent, \(H_{sim}\), for the DLFNN used by Han et al. and our RNN. The values \(H_{sim}\) have been binned into intervals 0.02 wide. The mean difference between \(H_{est}\) and \(H_{sim}\) was then calculated to produce the MAE. Both models have been trained on the same 10000 2D fBm tracks.}
    \label{fig:Dan model vs My model}
\end{figure}

The DLFNN used by Han et al. was trained on 1-dimensional fBm. However, the video PTM data collected here is 2-dimensional, so in order to match the model to the experiment, we have used 2D training data composed of two fBm simulations with the same length and value of \(H_{sim}\). This small change alone decreases the overall mean absolute error (MAE) of the DLFNN method by \(\sim20\%\) for a model with 15 steps. Switching to an RNN from the DLFNN brings an additional \(6\%\) improvement in MAE from 0.1060 to 0.0993. Figure \ref{fig:Dan model vs My model} show how the MAE varies with \(H_{sim}\) for the 2D DLFNN and RNN.
Before the tracks are fed into a NN, they must be normalised to stop step size becoming a factor in prediction. In the 2D case, the overall displacement, \(r_i = \sqrt{x_i^2+y_i^2}\), where \(x_i\) and \(y_i\) are the displacements of the two simulated tracks after \(i\) time intervals, is normalised between 0 and 1. The sequence of step sizes, \(r_{i+1}-r_i\), is then calculated and input to the NN.

When using a model with a single output neuron for short tracks, we found that the range of \(H_{est}\) was truncated close to \(H_{est}=0\) and to a lesser extent \(H_{est}=1\). Similar artefacts are observed using DLFNNs and are responsible for the increase in MAE at low \(H_{sim}\) that can be seen in both models in figure \ref{fig:Dan model vs My model}. To avoid this, we created another model with 21 output neurons corresponding to discrete increments of \(H\) 0.05 apart and used the two in tandem. The output activation function for this model is a softmax function that gives each neuron a normalised value between 0 and 1, interpreted as a probability that the given neuron represents the closest value of \(H\) to the true result. If the average response of all of the output neurons is taken, a response very similar to that of the single output neuron case is recovered. This suggests that the over/under prediction at low and high \(H\) stems from trying to recreate the full range of \(H\) behaviour using a single weight and bias on the output neuron. A simple algorithm was then used to find \(H_{est}\): if \(H_{1n}>0.2\), use \(H_{1n}\); if \(H_{21n}<0.2\), use \(H_{21n}\); otherwise, use \(H_{1n}\). \(H_{1n}\) and \(H_{21n}\) correspond to the prediction by a model with 1 and 21 output neurons respectively. The PC used to train the model also has a small impact on the outcome, so in order to properly compare models, they must be trained on the same computer and with the same training data. A more detailed discussion of the model architecture and computer can be found in the appendix section \ref{sec:NN architecture}.\\
\begin{figure*}
\centering
    \subfloat[\label{fig:Model performance:1N distribution}]{\includegraphics[width=0.9\columnwidth]{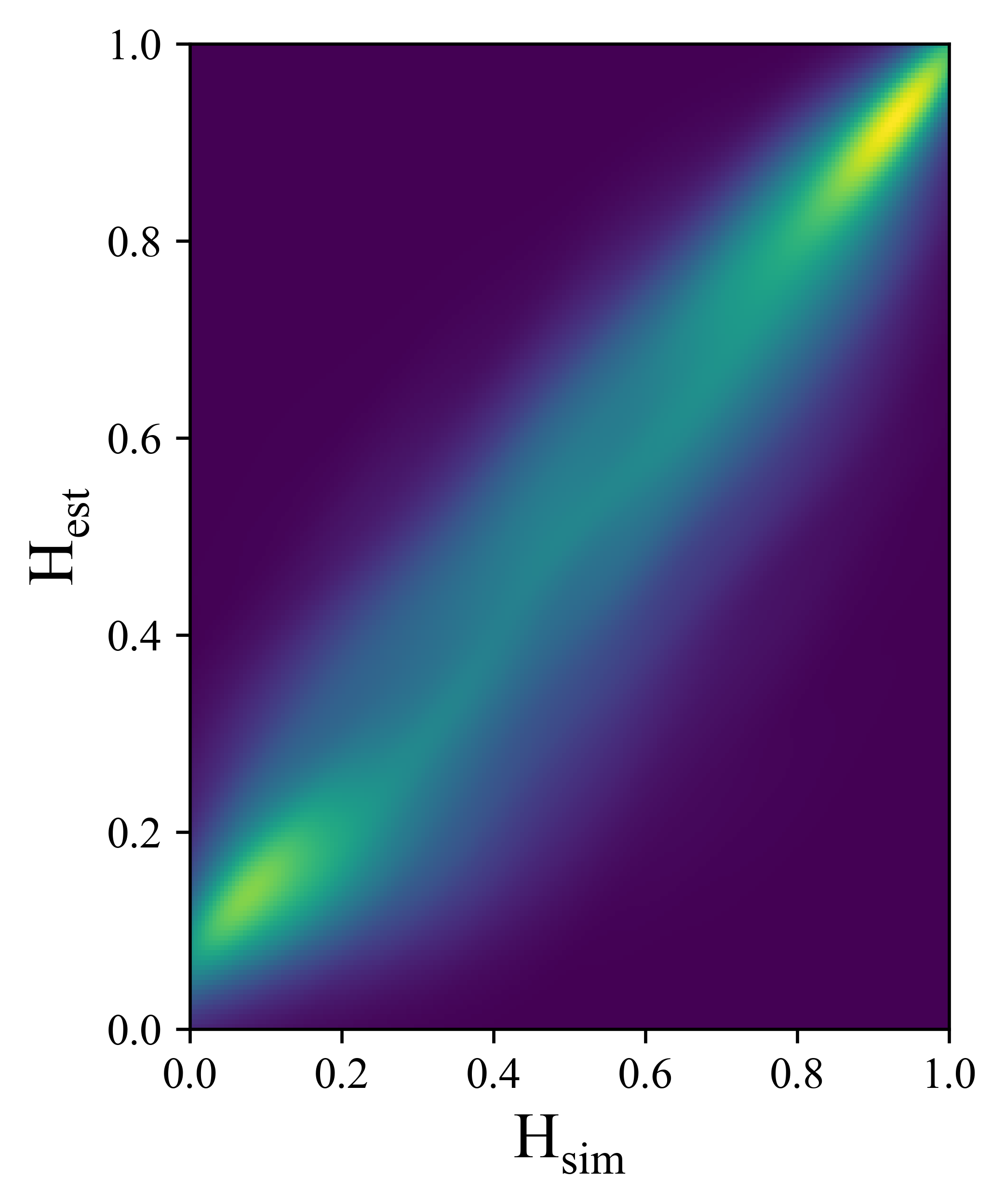}}    
    \subfloat[\label{fig:Model performance:Comb distribution}]{\includegraphics[width=0.9\columnwidth]{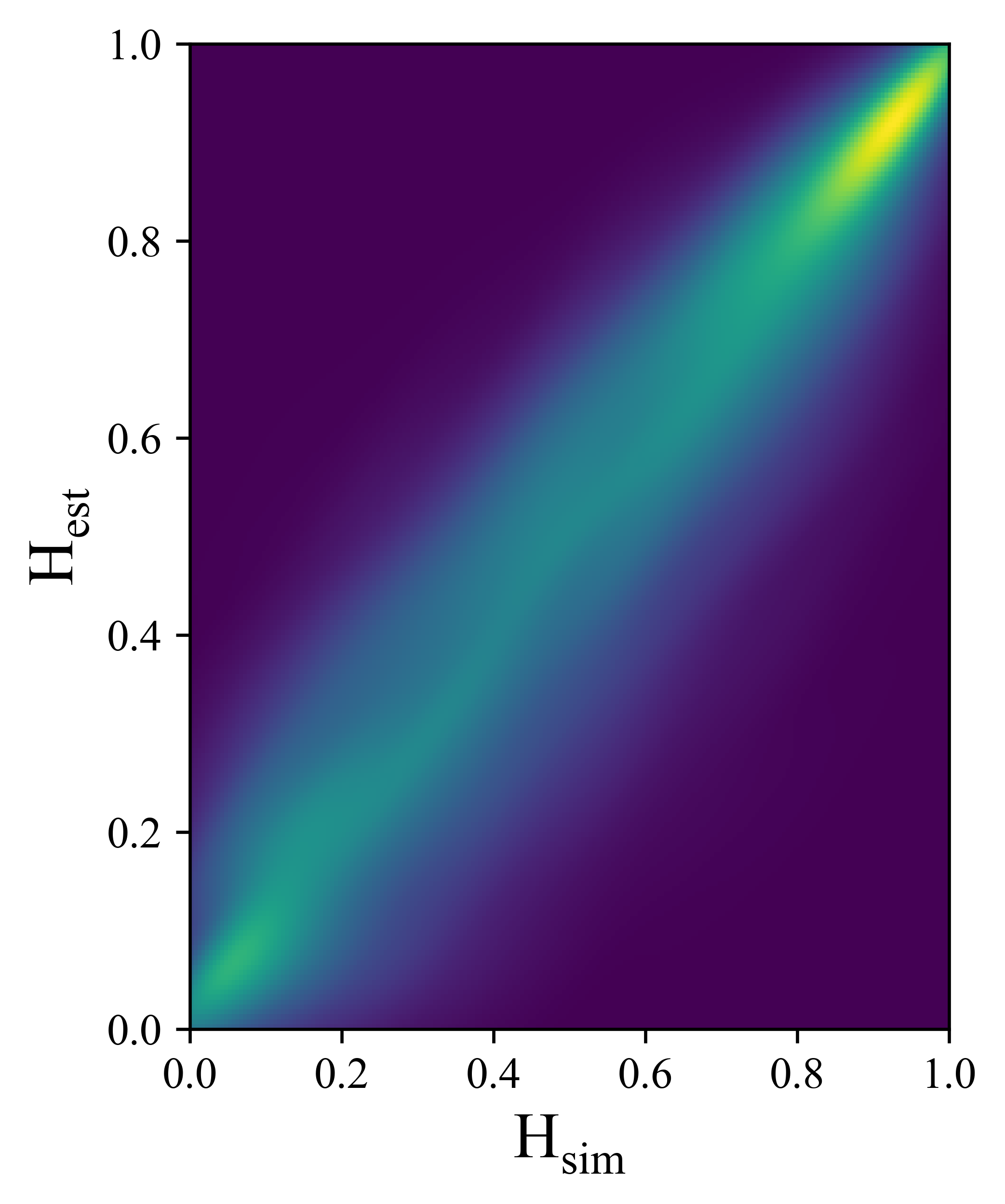}}
    
    \subfloat[\label{fig:Model performance:MAE plot}]{\includegraphics[width=0.9\columnwidth]{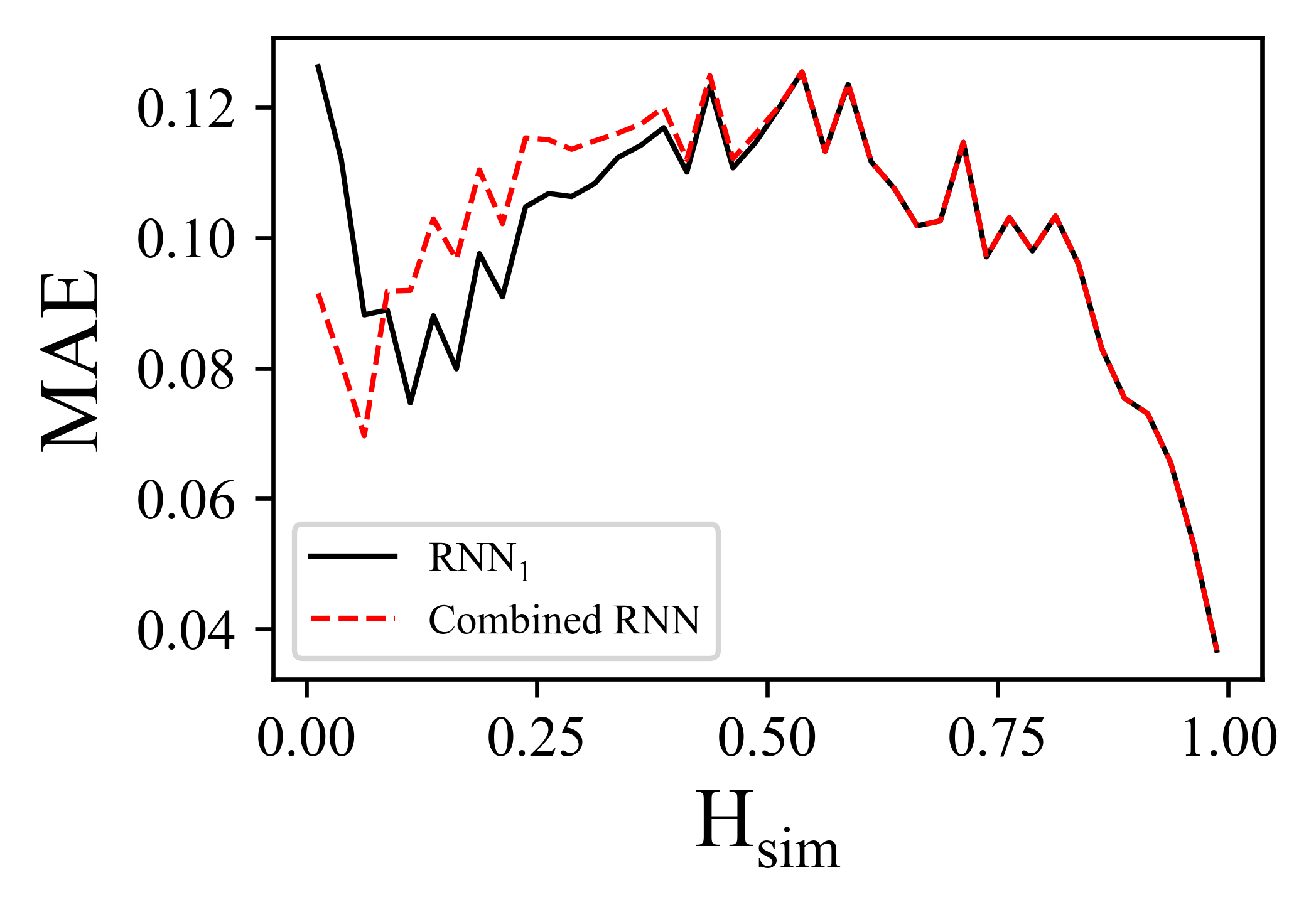}}
    \caption{The Gaussian KDE for the response of the one-neuron and combined models to the same 10000, 15-step fBm tracks with random \(H_{sim}\) are (a) and (b) respectively. The colour represents the probability of \(H_{est}\) being predicted at a given \(H_{sim}\). The mean absolute error, MAE, is plotted as a function of the simulated Hurst exponent, \(H_{sim}\), for each method in (c). }
    \label{fig:Model performance}
\end{figure*}
Gaussian kernel density estimate (KDE) plots for each method can be found in figures \ref{fig:Model performance:1N distribution} and \ref{fig:Model performance:Comb distribution}. They were made using the same 10,000 15-step fBm tracks with random values \(H_{sim}\). The colour on the plot represents the probability of getting a value of \(H_{est}\) for a given value of \(H_{sim}\).
A gap can clearly be seen at low \(H_{est}\) in the one-neuron case (figure \ref{fig:Model performance:1N distribution}) corresponding to the overshoot previously mentioned. The use of the 21-neuron model for low \(H\) estimation (figure \ref{fig:Model performance:Comb distribution}) has filled in this gap, removing the floor for \(H_{est}\). A comparison of the MAE for each method can be seen in figure \ref{fig:Model performance:MAE plot}. The overall MAE increases slightly to 0.1003 for the combined method. This is mostly due to the artificially low error at \(H_{sim}\approx0.2\) in the one output-neuron model, originating from the bunching of the values of \(H_{est}\), increasing the likelihood of a close prediction in that region.

\section{Results and discussion}
\subsection{Simulations}
\subsubsection{Hurst exponent}
If a real track moves with fBm, it should be possible to use the distribution of \(H_{est}\) to find the true value of \(H\), even if there is a systematic error associated with our model. To provide a measure, fBm tracks with 10,000 steps and single Hurst exponent, \(H_{sim}\), were simulated and fed into our NN. Given the non-linearity inherent in NN methods and the bounds on \(H\), it is not clear what type of distribution \(H_{est}\) will take. The simplest measures of the shape of a distribution are the first four moments about the mean: mean \((\mu)\), variance \((\sigma^2)\), skewness (\(\mu_3\)), and kurtosis \((\mu_4)\). Because the distributions are non-symmetric in general, the mean won't necessarily correspond to the peak position. A Gaussian kernel density estimate (KDE) with a bandwidth of 0.075 was therefore used to fit a general non-parametric probability density function (PDF) to each distribution, from which the peak could be estimated. A Gaussian KDE takes the sum of identical Gaussian distributions with centres at each value in the real distribution to create an estimate for the underlying PDF. This has the benefit of requiring no assumptions about the shape of the distribution. However, as Gaussians have no bounds, and \(0<H<1\), the resulting estimate is not a true PDF. As such, the KDEs used here have been cut off at 0 and 1 and re-normalised so that the area under the curve is still equal to 1. This has the effect of creating a discontinuity at the extremes for some distributions, as can be particularly seen in figure \ref{fig:H hists:Track 2 H hist}. Even still, the KDEs provide a good estimate of the peaks in the distributions. 

\begin{figure*}
\centering
    \subfloat[\label{fig:H simulations:H sim mean}]{\includegraphics[width=0.9\columnwidth]{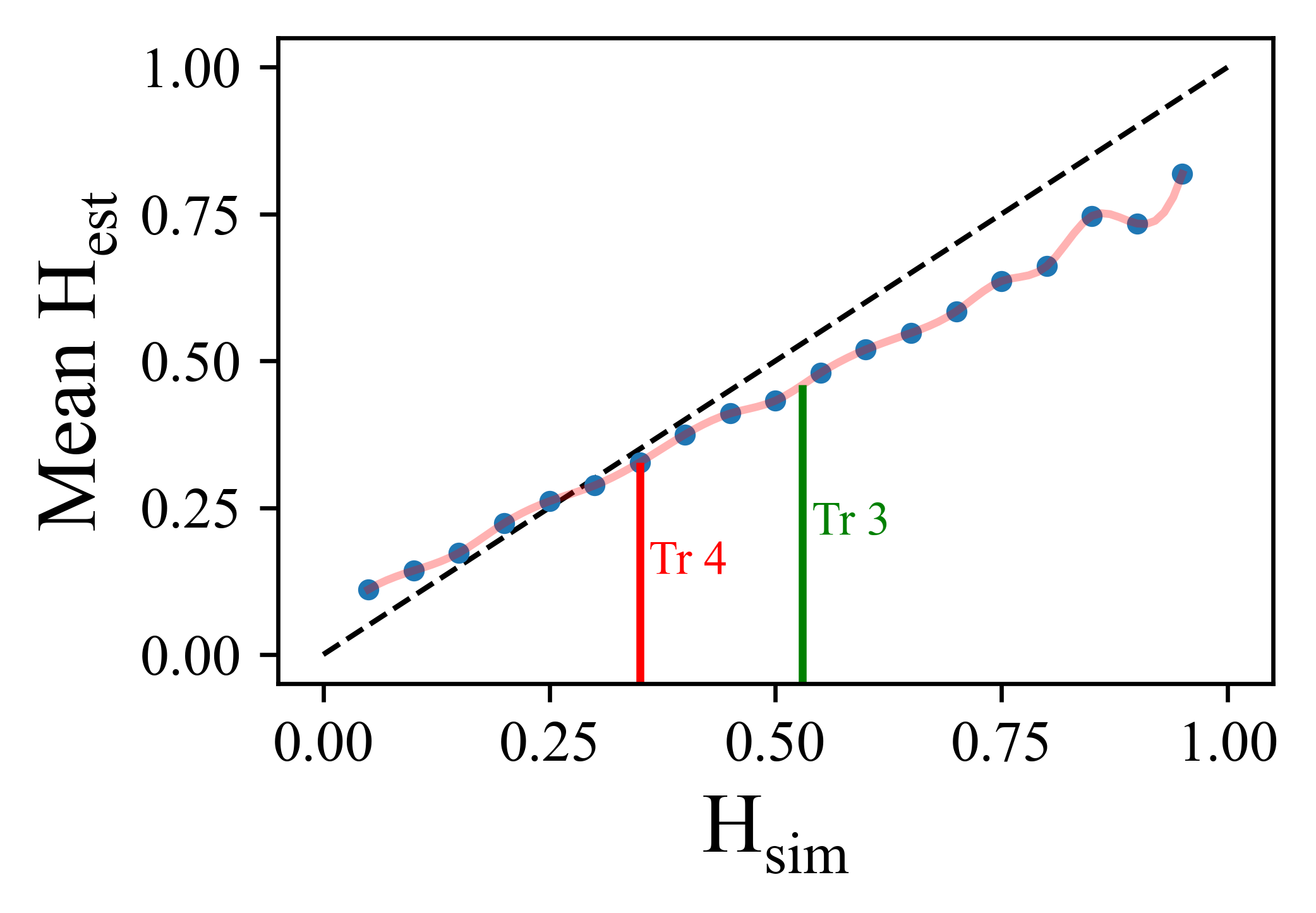}}
    \subfloat[\label{fig:H simulations:H sim skew}]{\includegraphics[width=0.9\columnwidth]{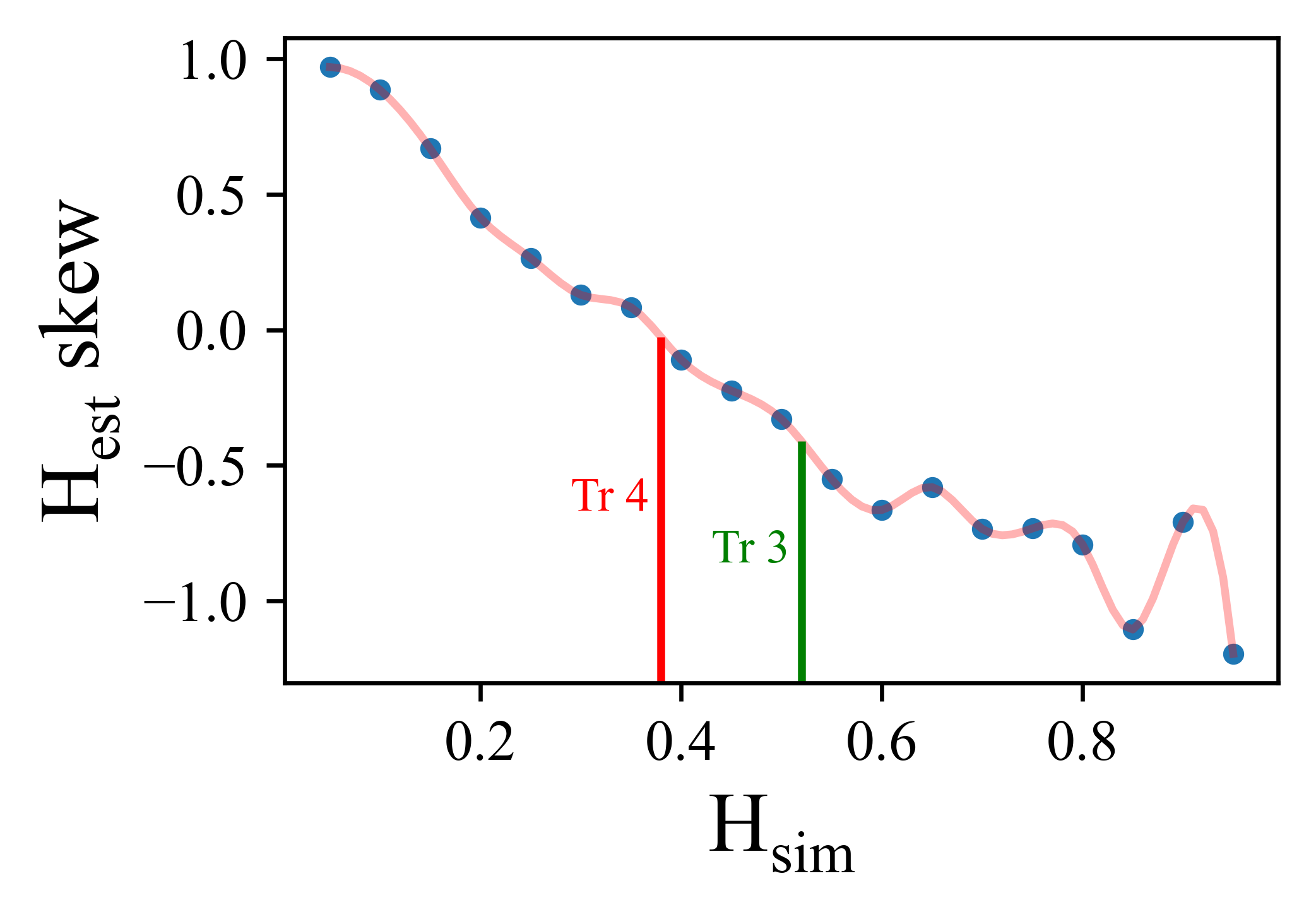}}

    \subfloat[\label{fig:H simulations:H sim KDE}]{\includegraphics[width=0.9\columnwidth]{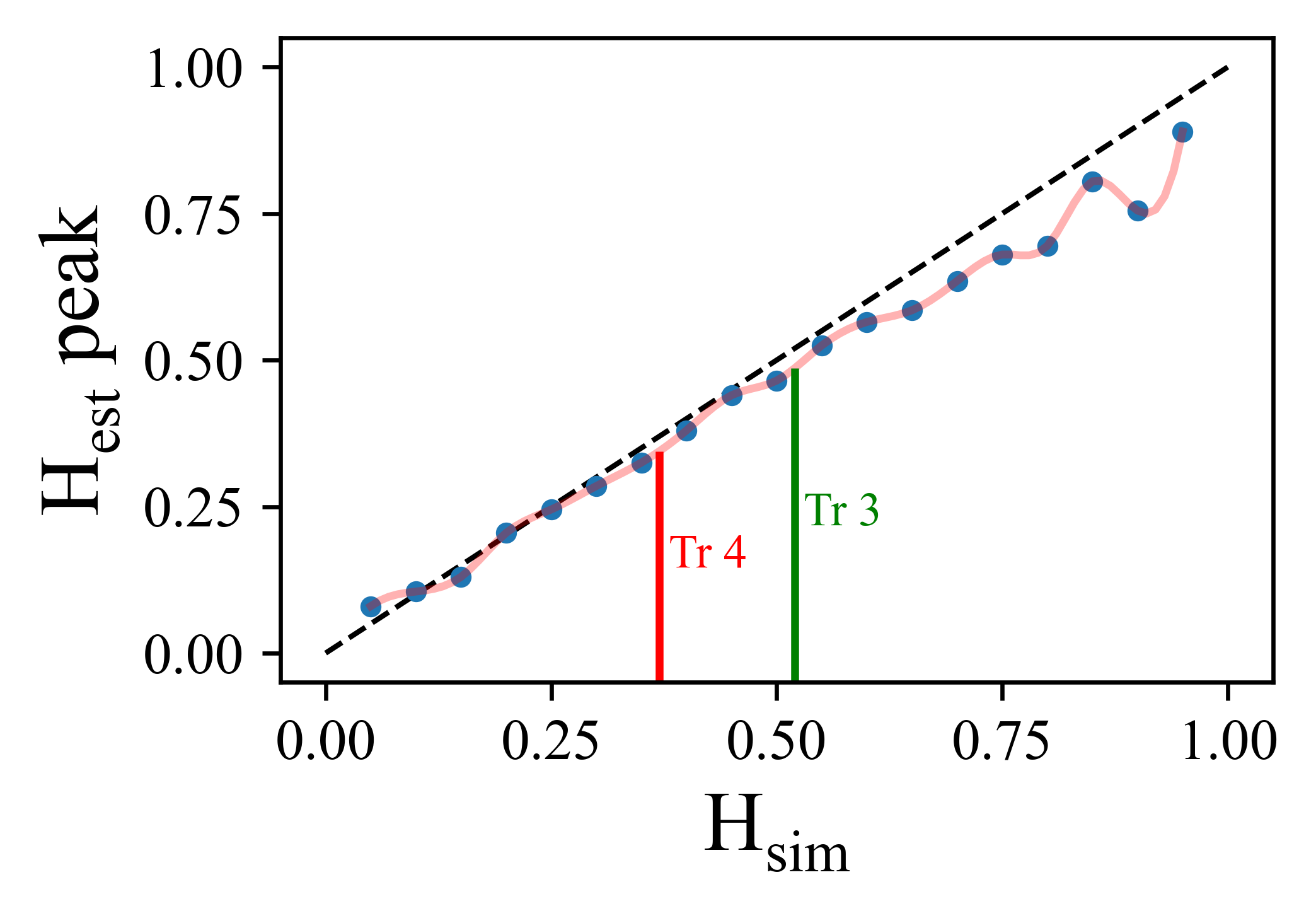}}
    
    \caption{(a) - (c) show how the mean, skew, and peak position respectively of a KDE fit with bandwidth 0.075, of the distributions of $H_{est}$ vary for simulated fBm random walks with 10000 steps. A cubic spline has been used to interpolate between the data points and the values of \(H_{est}\) suggested for tracks 3 and 4 by their distributions have been plotted on each figure in green and red respectively. These results are summarised in table \ref{tab:H sim estimates}. Black dashed lines following \(mean H_{est}=H_{sim}\) have been included in (a) and (c) to show where an ideal prediction would be.}
    \label{fig:H simulations}
\end{figure*}

The mean, skewness, and KDE peaks plotted as a function of \(H_{sim}\) can be found in figure \ref{fig:H simulations}. The plots for the variance and kurtosis have been excluded as they have complex, non-monotonic shapes including plateaus and turning points. This means that a single value of \(\sigma^2\) or \(\mu_4\) could correspond to multiple values of \(H\) or give \(H\) with low accuracy. Given the non-linear nature of the NN, it is perhaps unsurprising that shape of the distributions produced do not vary smoothly with \(H_{sim}\). Looking at the plot of the mean of the distribution of \(H_{est}\) in figure \ref{fig:H simulations:H sim mean}, there are small biases in the prediction stemming from the asymmetry of the distributions. The model seems to work best for \(H_{sim}<0.5\) as it has been fine-tuned to perform best in the subdiffusive regime as this is where we expect the majority of motion to be in our LGN. This is mirrored in the skew plot (figure \ref{fig:H simulations:H sim skew}) where the curve is at its smoothest for \(H_{sim}<0.5\). The plot of KDE peak in figure \ref{fig:H simulations:H sim KDE} again shows the best performance for \(H_{sim}<0.5\) and in general produces more accurate results than the mean. In order to use these plots as a tool for estimating \(H\) for real tracks, a cubic spline has been used to interpolate between data points. The results of this can be seen in table \ref{tab:H sim estimates} for tracks with unimodal distributions of \(H_{est}\).

\subsubsection{Anisotropy}
In the context of fBm, it makes intuitive sense that the anisotropy of a particle's track should be connected to the value of \(H\). If a particle's motion is persistent, it will look more like a straight line the larger \(H\) becomes. Conversely, anti-persistence will tend to create more symmetrically distributed tracks that change direction often. In order to test this hypothesis, simulated tracks with a single value of \(H_{sim}\) and 500,000 steps were split into sections 45 steps long. Sections 45 steps long were chosen so as to balance the increasing statistical fluctuation inherent in stochastic tracks with fewer steps and the ability to see local dynamics and provide comparison with the NN estimations of \(H\). A track, sampled \(N\) times to give a set of position vectors, can be thought of as analogous to a distribution of \(N\) identical particles located at each sampled position. This allows the anisotropy to be calculated with the principle moments of the 2D gyration tensor of the track section. The gyration tensor of a distribution of \(N\) steps, \(g_{mn}\) is given by,
\begin{equation}\label{eq:GT}
    g_{mn} = \frac{1}{N}\sum_{i=1}^Nr^{i}_mr^{i}_n,
\end{equation}
where \(r^{i}_m\) is the m\(^{th}\) spatial coordinate relative to the centre of mass of the i\(^{th}\)  step. The principle moments of \(g_{mn}\) are then given by the eigenvalues of the tensor. Working in 2D, there will be two eigenvalues, \(\lambda_1\) and \(\lambda_2\). A circularly symmetric distribution of particles will yield \(\lambda_1=\lambda_2\), while a distribution along a line will result in \(\lambda_1=0\) (if the coordinate system is chosen such that \(\lambda_1\leq\lambda_2\)). The anisotropy, \(A\), can then be normalised to a value between 0 and 1 through the operation,
\begin{equation}\label{eq:Anisotropy}
    A=\frac{\lambda_2-\lambda_1}{\lambda_1+\lambda_2}.
\end{equation}
\begin{figure*}
\centering
    \subfloat[\label{fig:A simulations:A mean}]{\includegraphics[width=0.9\columnwidth]{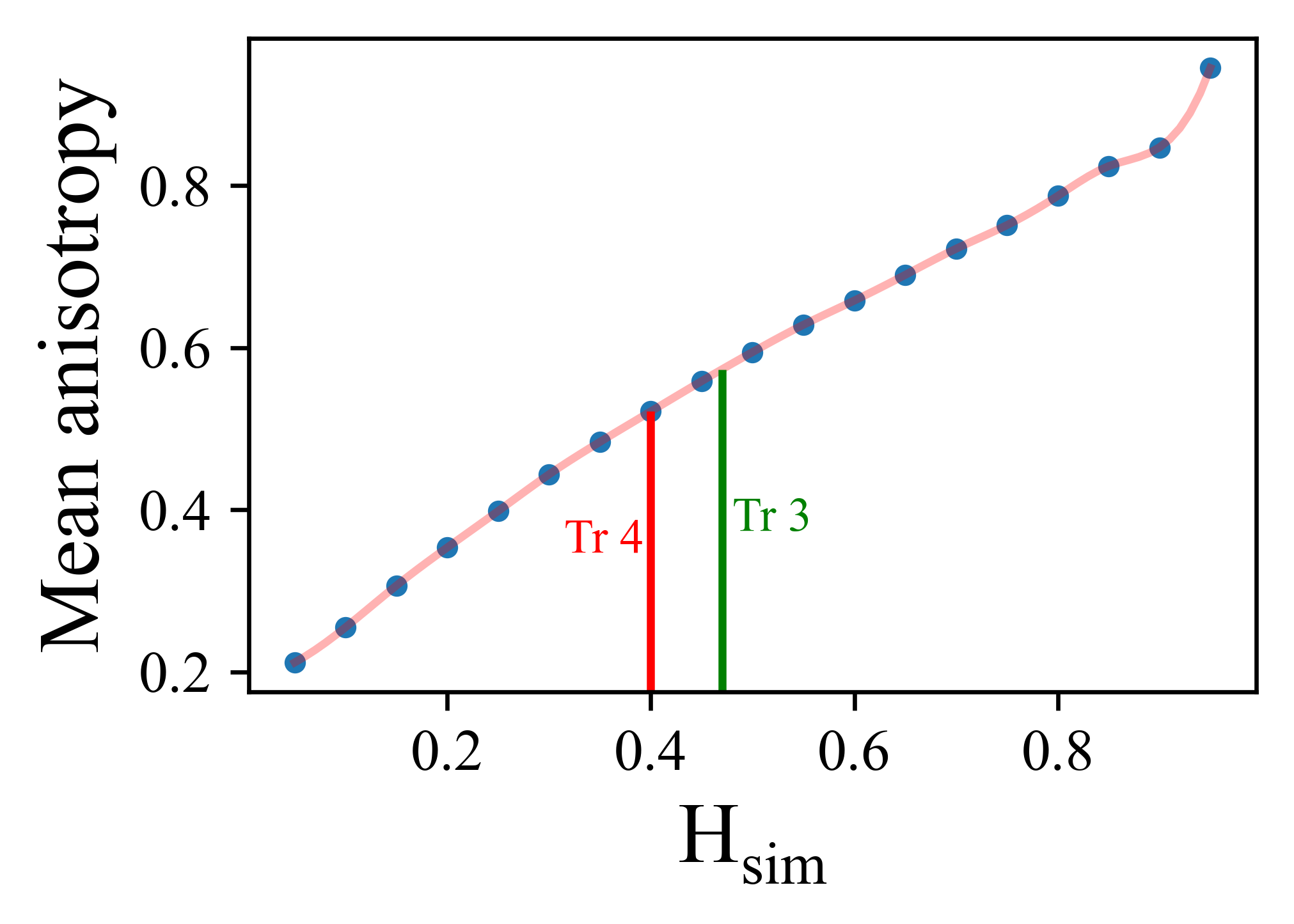}}
    \subfloat[\label{fig:A simulations:A skew}]{\includegraphics[width=0.9\columnwidth]{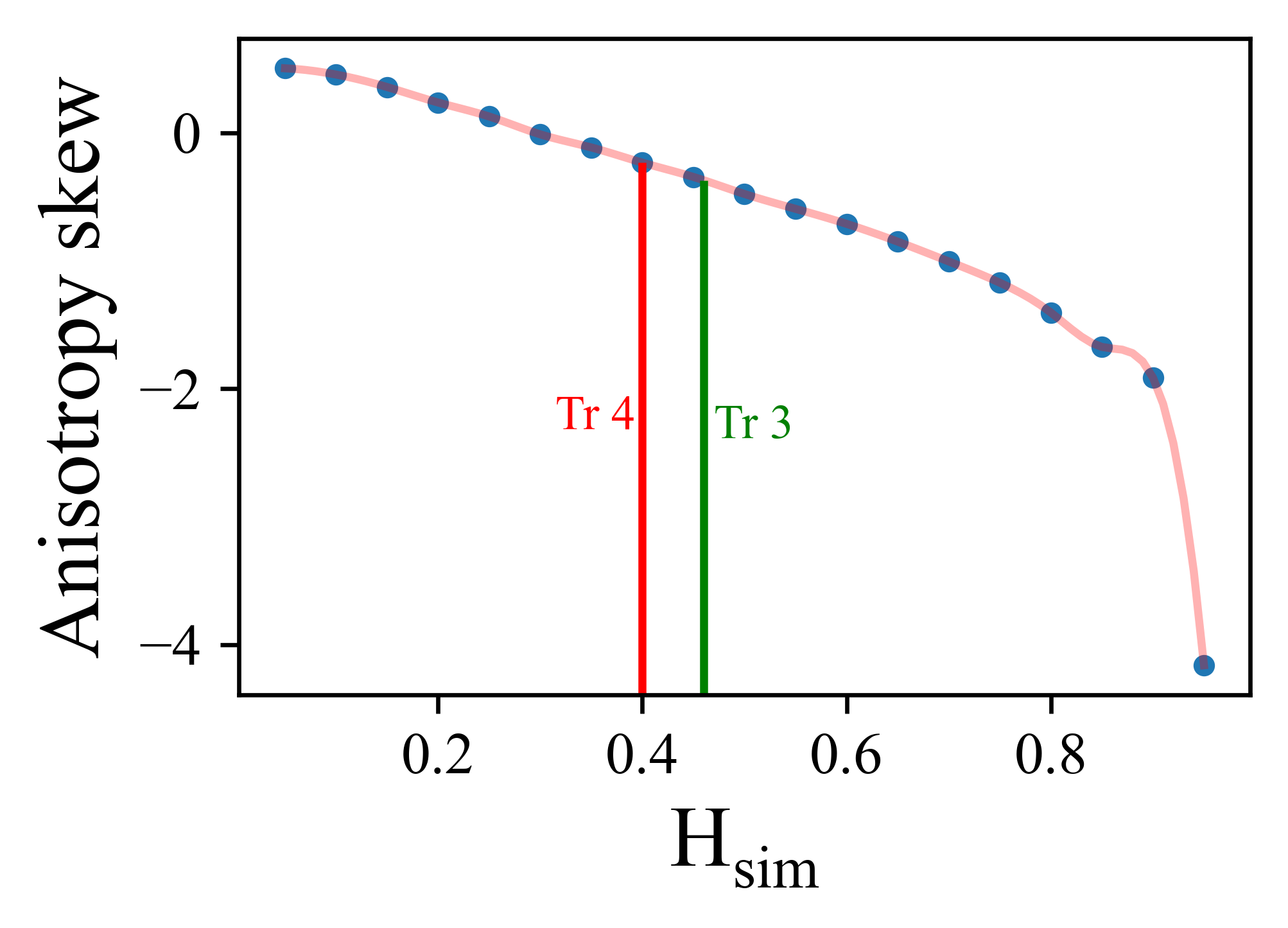}}
    
    \subfloat[\label{fig:A simulations:A KDE}]{\includegraphics[width=0.9\columnwidth]{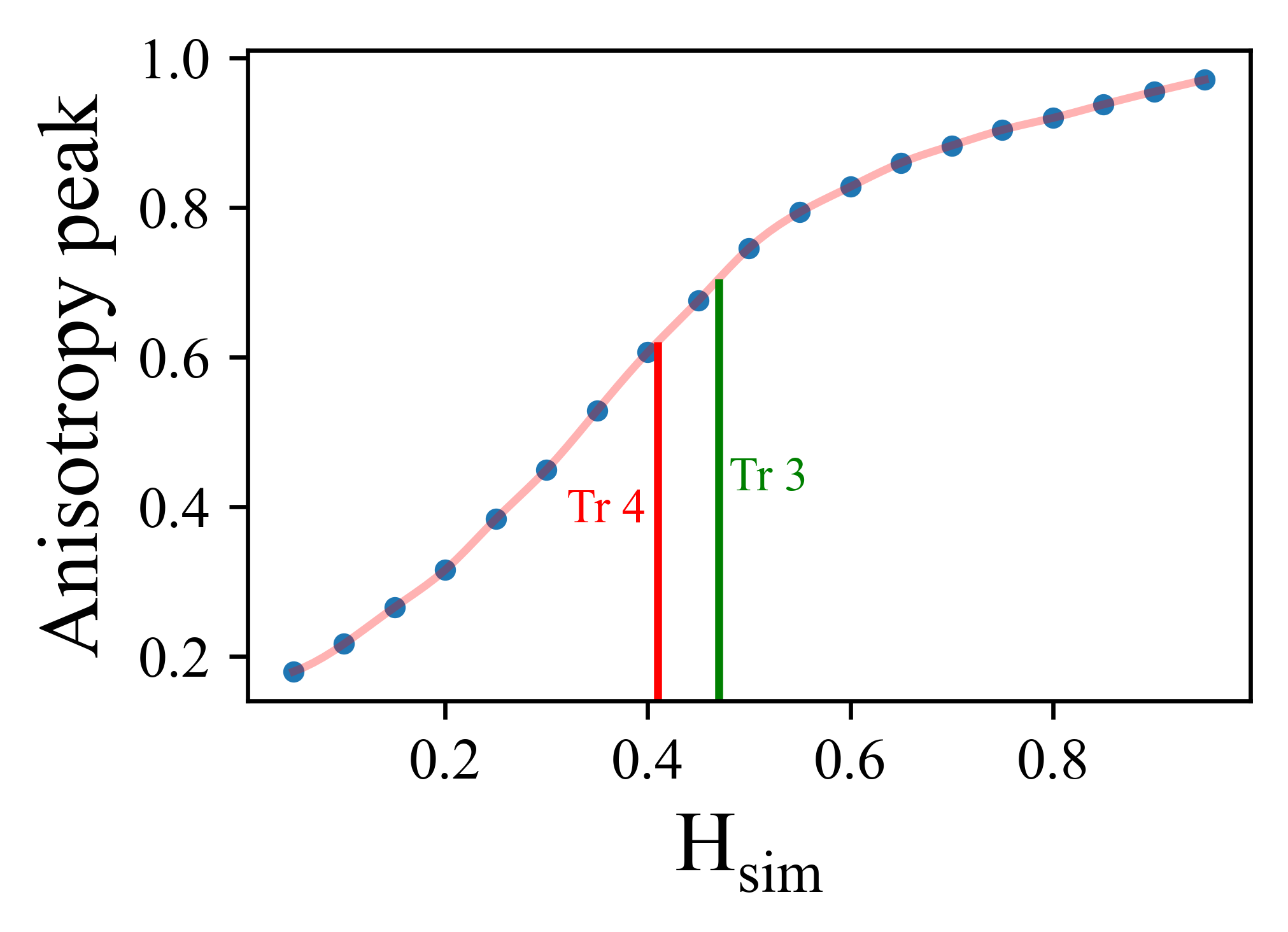}}
    
    \caption{(a) - (c) show how the mean, skew, and KDE peak of the distribution of anisotropy vary for simulated fBm random walks with different Hurst exponents. Each track has 500000 steps and a single value of H. The anisotropy is calculated in sliding windows of 45 steps using equation \ref{eq:Anisotropy}. The moments can then be calculated from the resulting distribution. A cubic spline has been used to interpolate between the data points and the values of H suggested by their distributions for tracks 3 and 4 have been plotted on each figure.}
    \label{fig:A simulations}
\end{figure*} 

Since \(0<A<1\), the same KDE method used for the distributions of \(H_{est}\) was again employed here, and plots analogous to those of figure \ref{fig:H simulations} for \(A\) can be seen in figure \ref{fig:A simulations}. Each moment varies smoothly with \(H_{sim}\), indicating that the anisotropy is indeed related to \(H\). However, the variance and kurtosis both include extended turning points or plateaus meaning estimations based on these moments may be inaccurate, so they have again been neglected. The mean, skewness, and peaks are all monotonic in the main region of interest of \(H_{sim}\lesssim0.5\), so should be useful in estimating \(H\). Cubic splines have been used in the same way as figure
\ref{fig:H simulations} to interpolate the data, aid the eye, and allow for anisotropy to be used to estimate \(H\) in real tracks.

\subsection{Hurst exponent and anisotropy}
\begin{figure*}
\centering
    \subfloat[\label{fig:H tracks:Track 1 H track}]{\includegraphics[width=0.9\columnwidth]{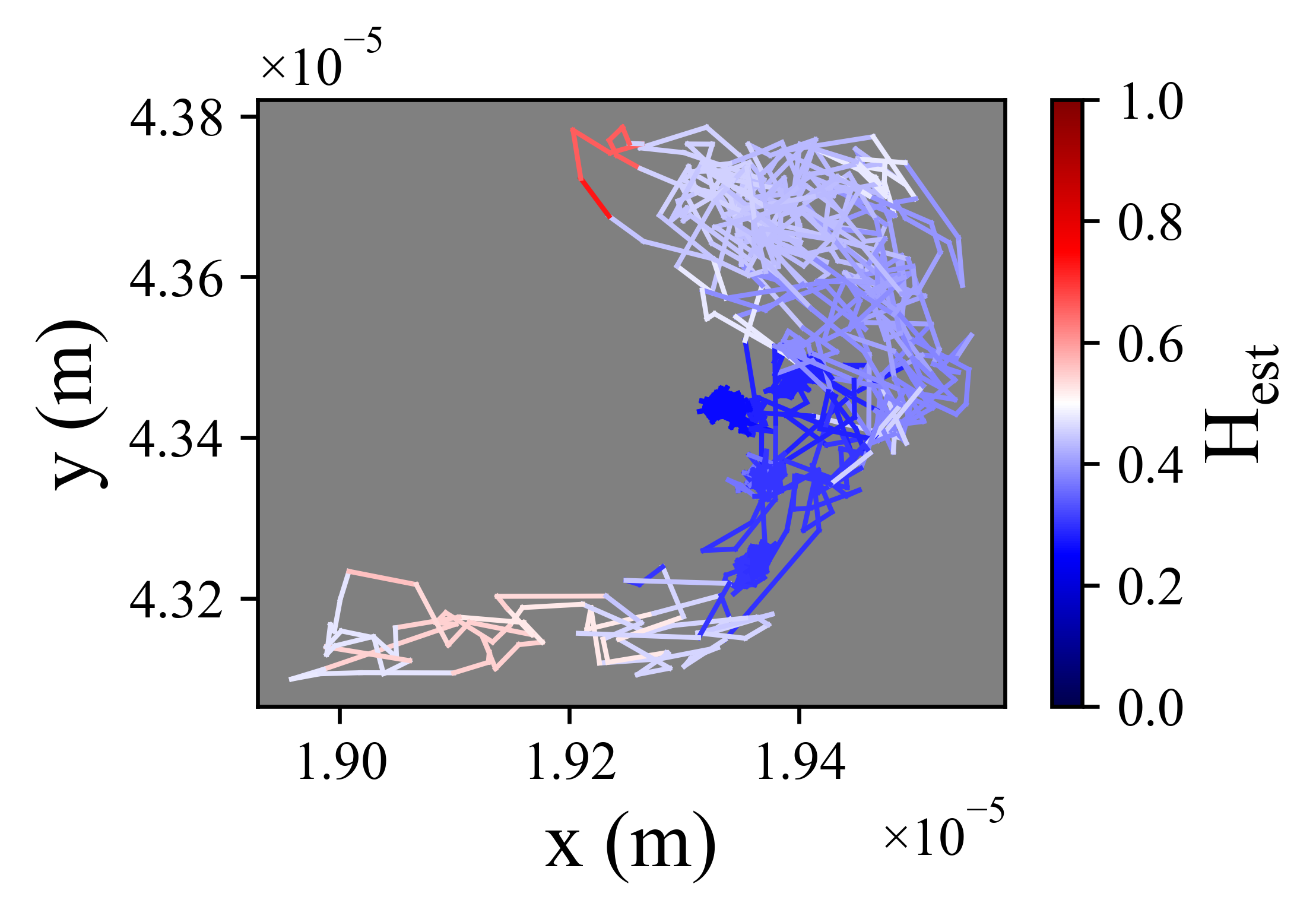}}
    \subfloat[\label{fig:H tracks:Track 2 H track}]{\includegraphics[width=0.9\columnwidth]{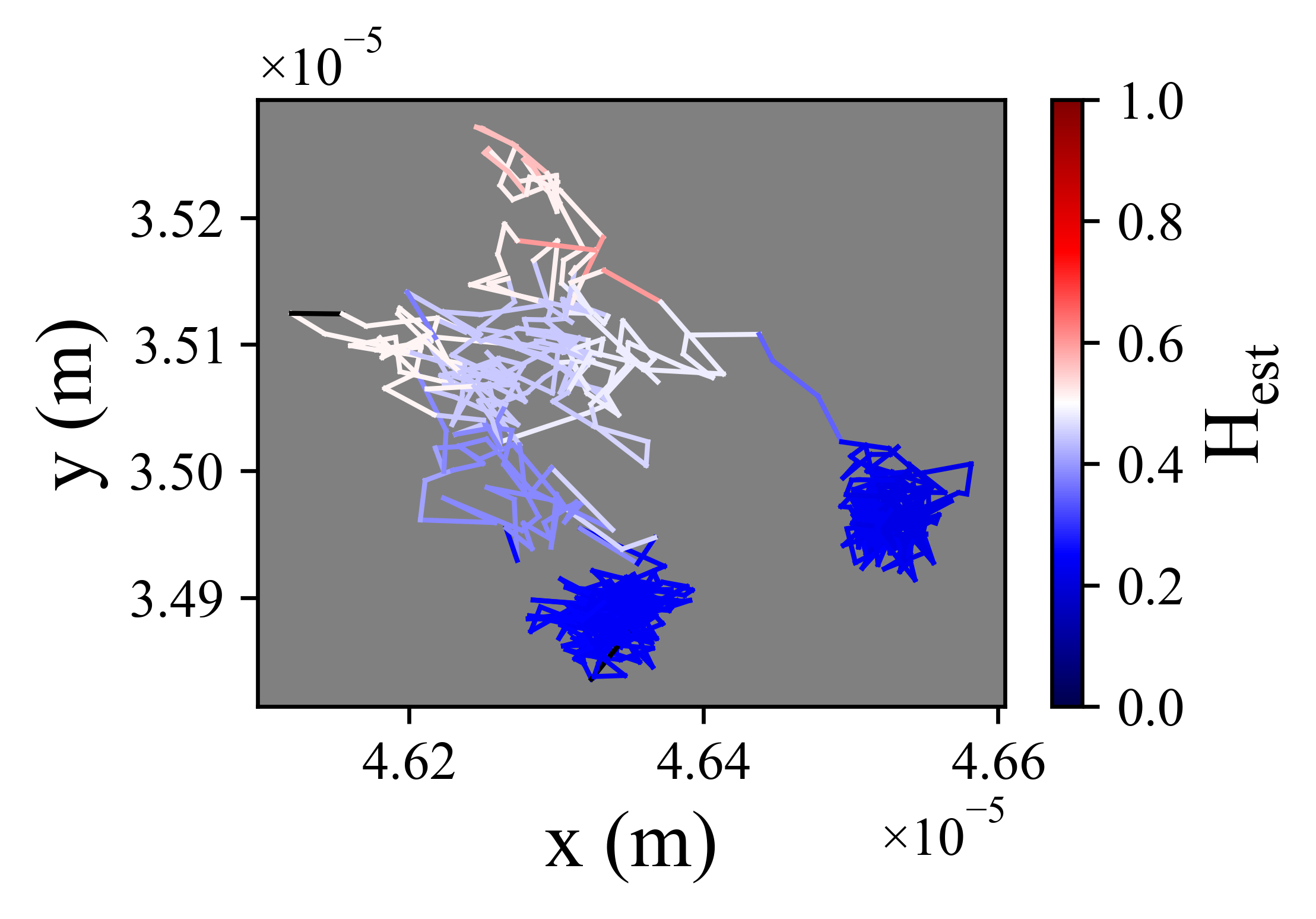}}
    
    \subfloat[\label{fig:H tracks:Track 3 H track}]{\includegraphics[width=0.9\columnwidth]{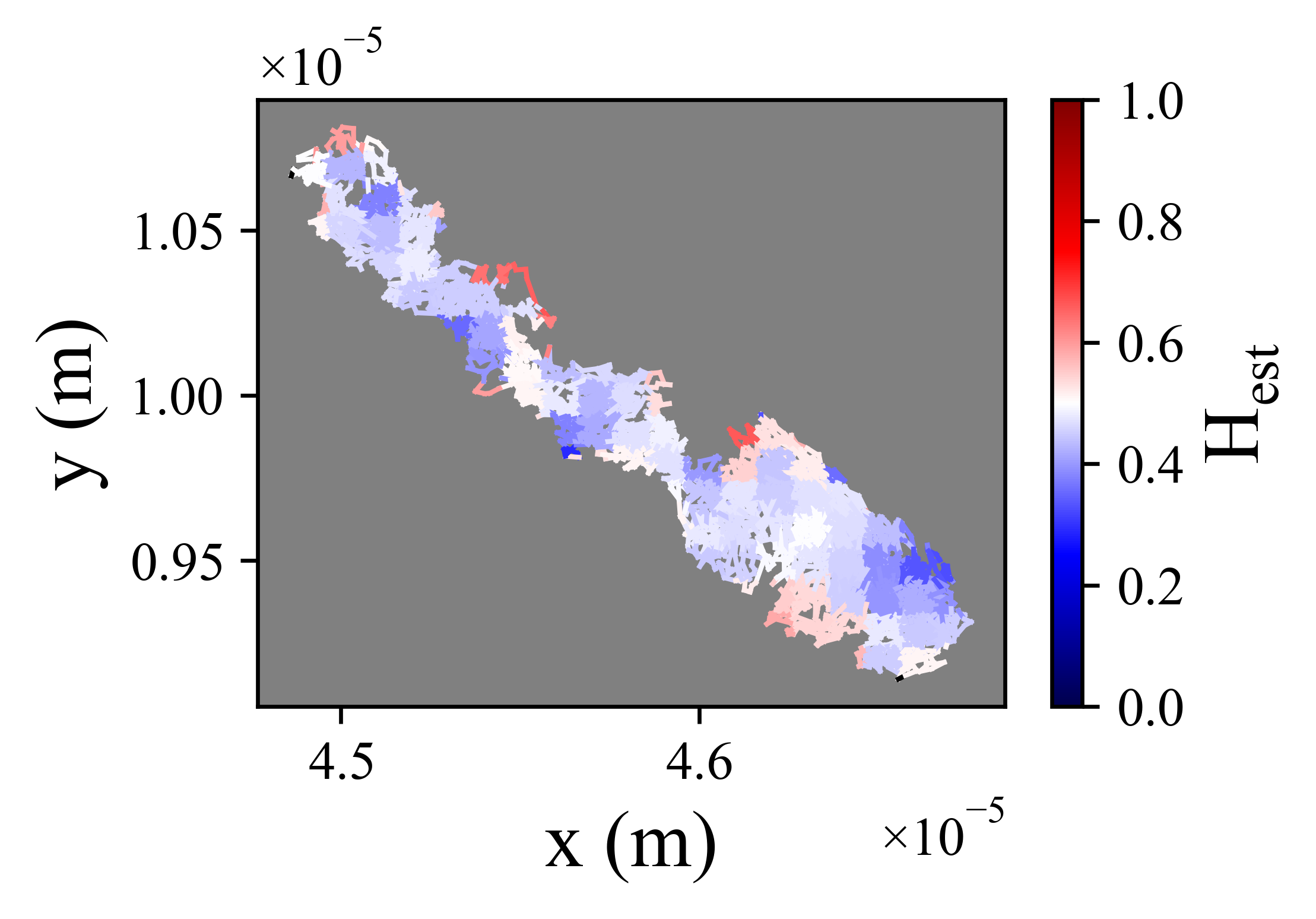}}
    \subfloat[\label{fig:H tracks:Track 4 H track}]{\includegraphics[width=0.9\columnwidth]{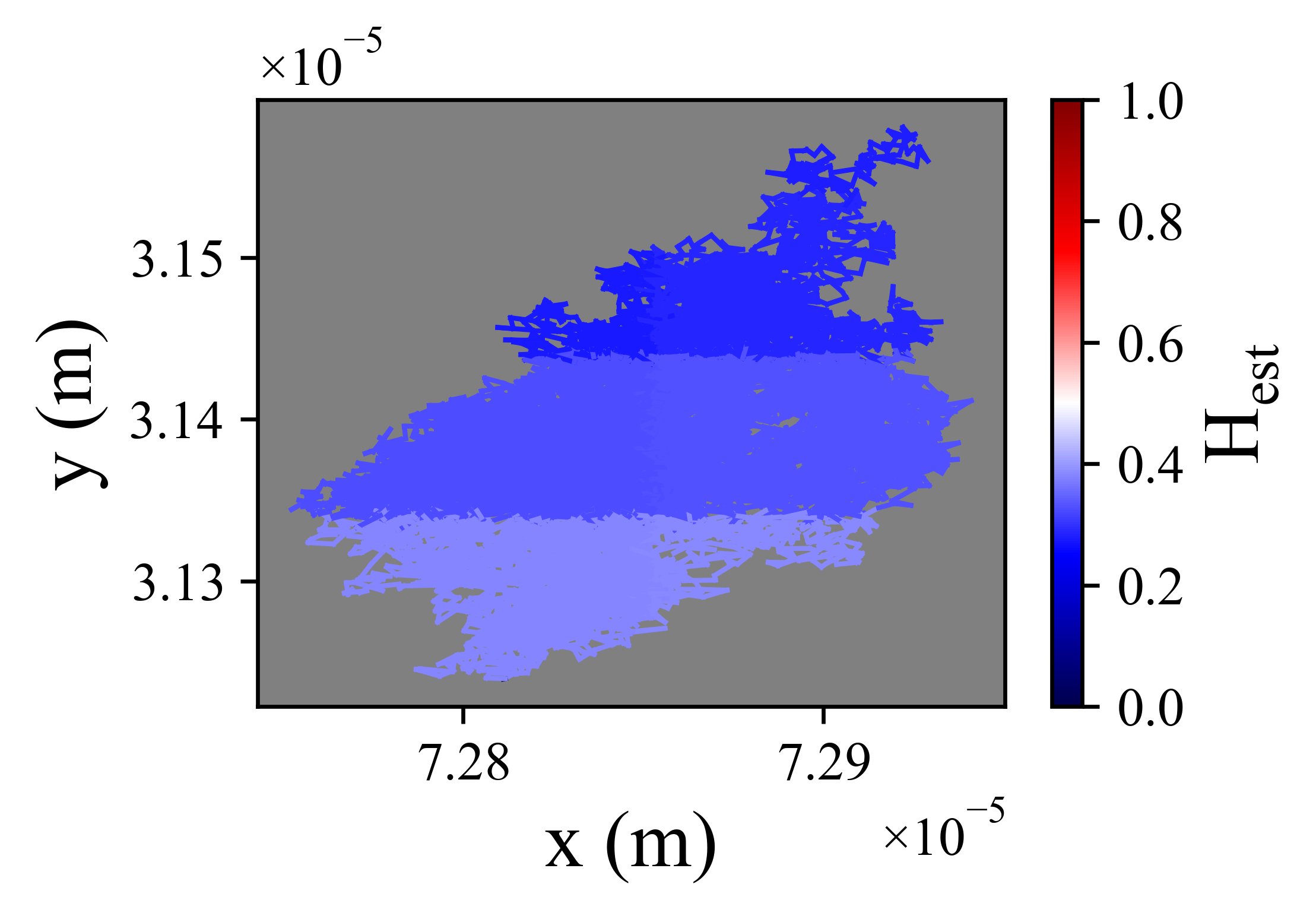}}
    
    \caption{(a), (b), (c), and (d) show tracks 1, 2, 3, and 4 respectively with the Hurst exponent mapped. H has been estimated in 15-step sections of each track and then spatially averaged into \(0.1\; \mu m\) boxes. Tracks 1 and 2 show clear examples of caging (low \(H_{est}\), sub-diffusive regions) where the particle spends the majority of its time. These are surrounded by areas where the particle moves more freely, even becoming super-diffusive. Track 3 and 4 show more uniform behaviour, 3 moving mostly diffusively and 4 subdiffusively.}
    \label{fig:H tracks}
\end{figure*}

Four tracks that highlight the diversity of behaviour present in our LGN samples were chosen, labelled as tracks 1-4. Track 1 (T1) was obtained from a data set taken at 50 fps, tracks 2 (T2) and 3 (T3) were taken at 125 fps and track 4 (T4) at 5000 fps. The tracks with \(H_{est}\) overlaid can be seen in figure \ref{fig:H tracks}. The central point of of each 15-step sliding window has been assigned the value of \(H_{est}\) output by our RNN. Each track has then been split into \(0.1 \;\mu m\) squares and the mean value of \(H_{est}\) has been calculated in each square and mapped accordingly.

\begin{figure*}
\centering
    \subfloat[\label{fig:H hists:Track 1 H hist}]{\includegraphics[width=0.9\columnwidth]{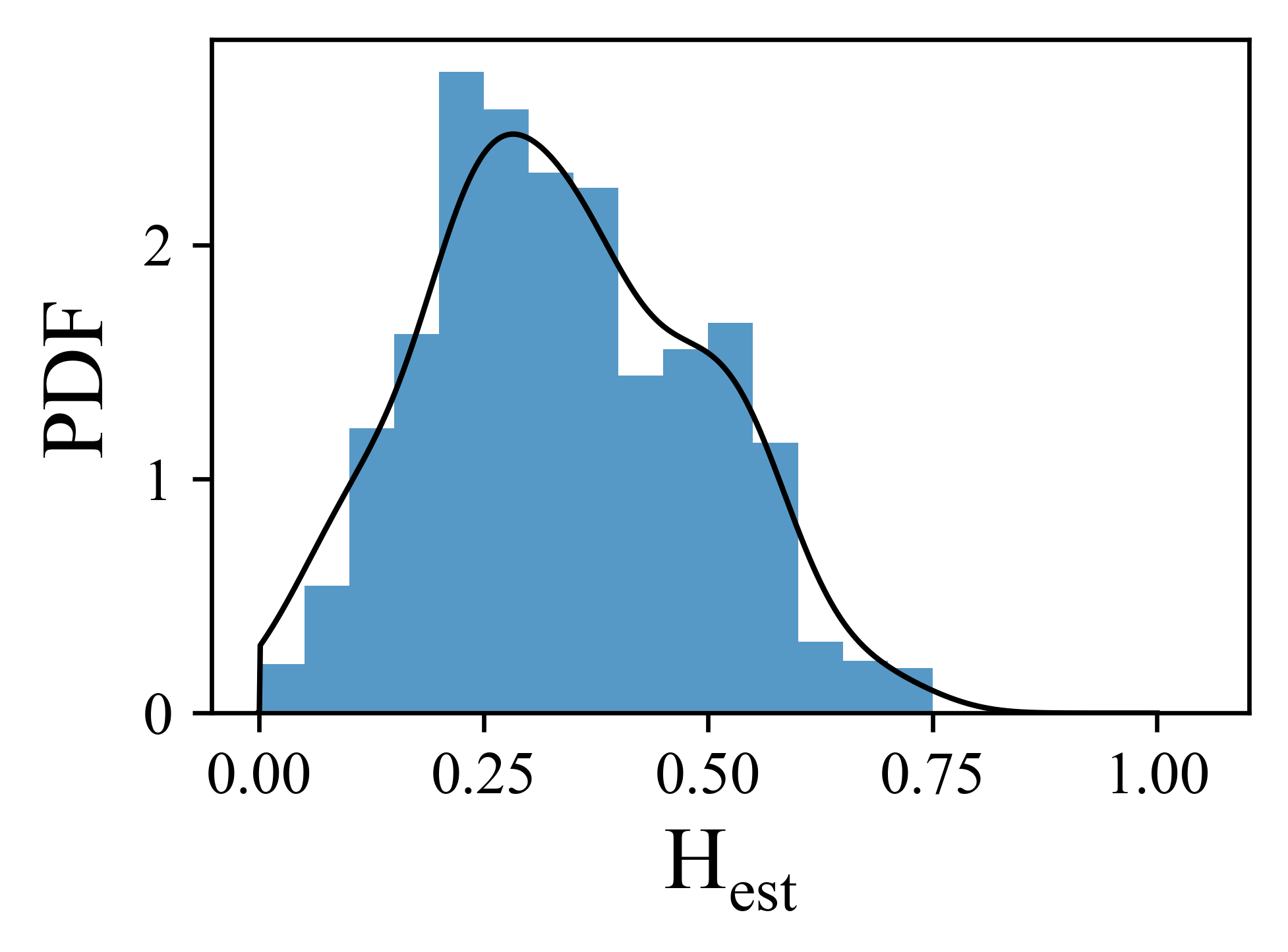}}
    \subfloat[\label{fig:H hists:Track 2 H hist}]{\includegraphics[width=0.9\columnwidth]{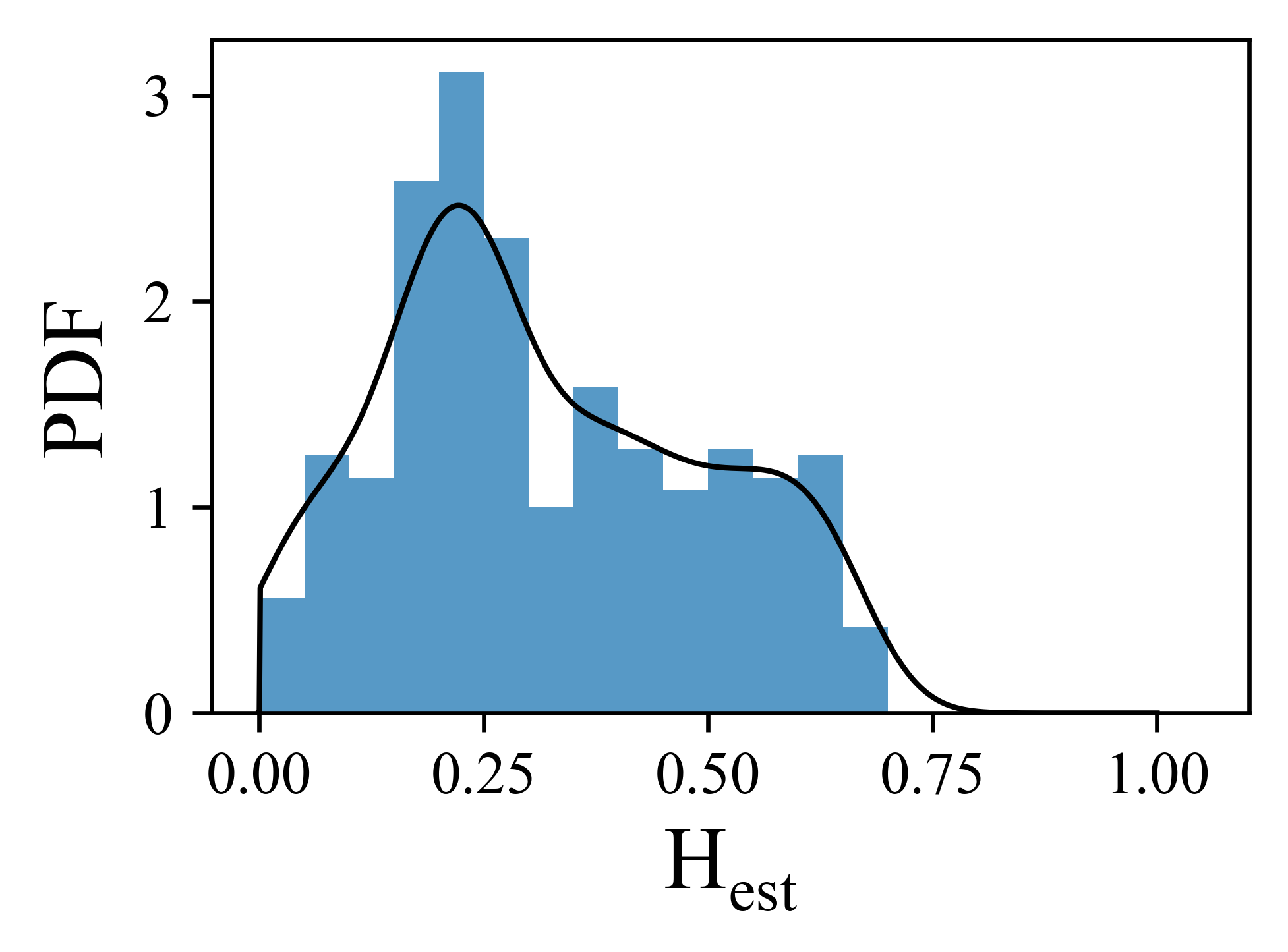}}
    
    \subfloat[\label{fig:H hists:Track 3 H hist}]{\includegraphics[width=0.9\columnwidth]{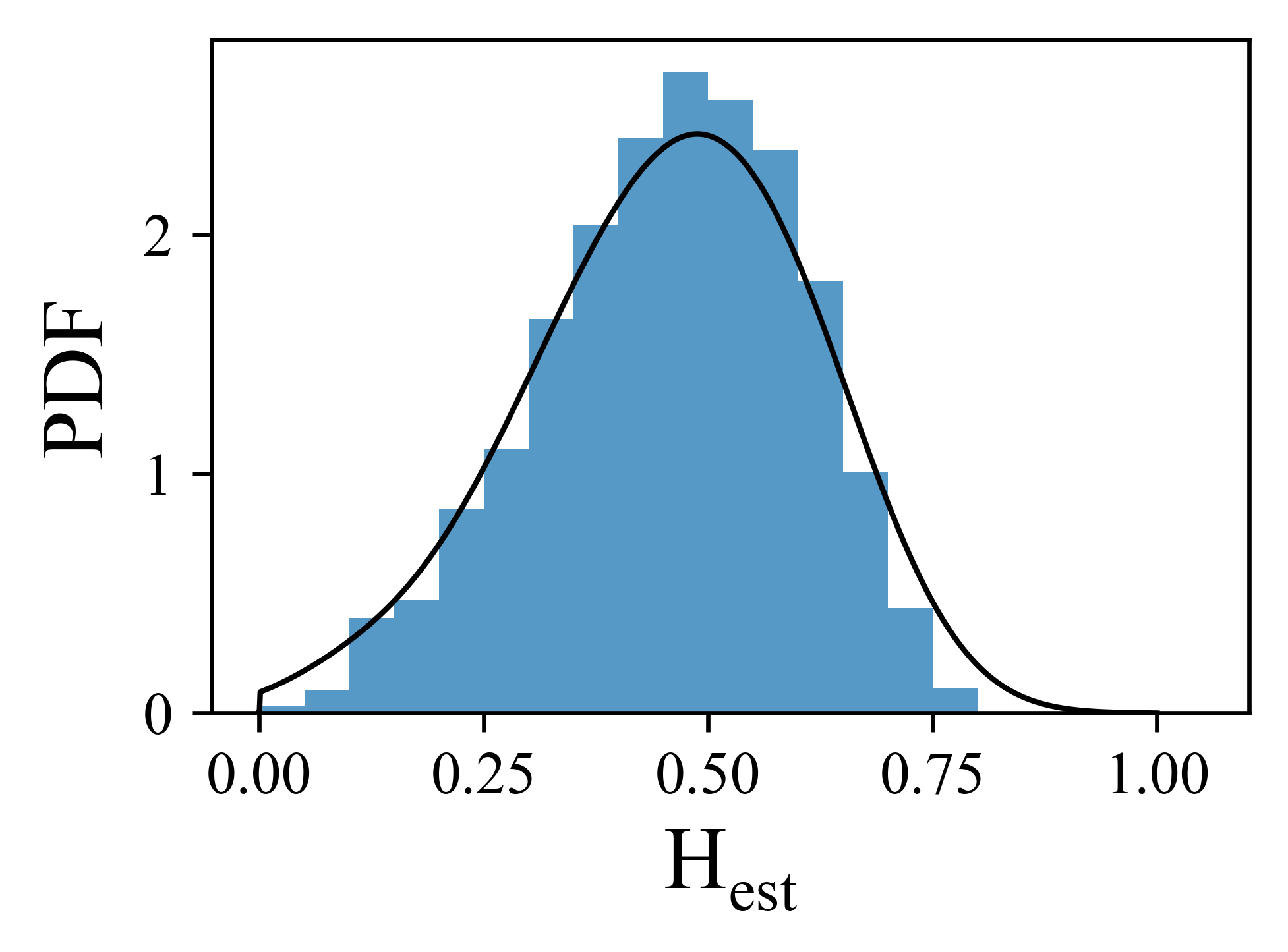}}
    \subfloat[\label{fig:H hists:Track 4 H hist}]{\includegraphics[width=0.9\columnwidth]{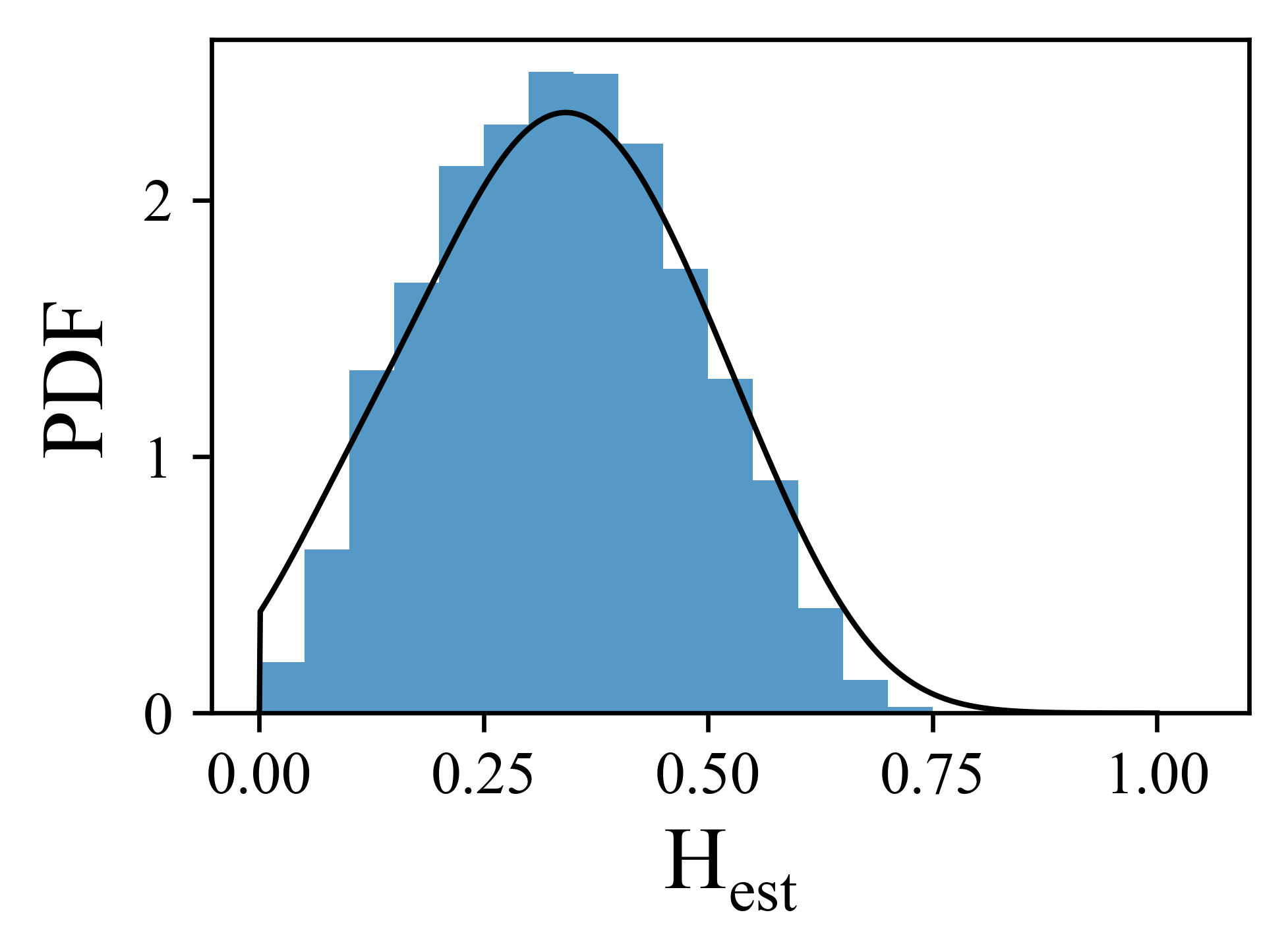}}

    \caption{(a)-(d) show the probability density functions (PDFs) for the distribution of the estimated Hurst exponent, \(H_{est}\), for tracks 1-4. The orange curve superposed on each figure is a Gaussian kernel density estimation (KDE) for the PDF. The spatial averages of this can be seen in figure \ref{fig:H tracks}. (a) and (b) are both multimodal distributions, while (c) and (d) are unimodal. The values of H for the primary peaks of the KDE for (a)-(d) are 0.285, 0.225, 0.490, and 0.340. The secondary peak in (a) comes at \(H\approx0.48\). The secondary peak in (b) is extended with a centre at at \(H\approx0.5\).}
    \label{fig:H hists}
\end{figure*}

T1 and T2 (figures \ref{fig:H tracks:Track 1 H track} and \ref{fig:H tracks:Track 2 H track}) show clear evidence of caging phenomena where the particle is trapped locally. In both cases, the cages account for a relatively small amount of the spatial extent of the track even though they are the regions in which the particle spends the majority of its time. The cages are on the order of the size of the particles (0.5 \(\mu\)m diameter) and they are interpreted by the neural network as subdiffusive as the particle's motion is tightly bounded leading to anti-persistent behaviour. Outside of these cages, the motion of the particles is mostly diffusive or subdiffusive with very local superdiffusive regions. We expect the superdiffusive regions are an expression of the stochastic nature of the motion, since on average there is no driving force for this type of motion. Caging can arise due to particle crowding in concentrated suspensions. The presence of both caging and extended, diffusive, colloidal motion in the same local regions suggests that this isn't the whole story here. There must be structures, be they multilamellar vesicles or lamellar domains, which trap the particles temporarily before they escape into the fluid, which then allows freer movement. Figure \ref{fig:H hists} shows probability density functions (PDFs) for the distributions of \(H_{est}\) for T1-4 with Gaussian KDEs fit to each distribution. The effect of the cages in T1 and T2 is reflected in figures \ref{fig:H hists:Track 1 H hist} and \ref{fig:H hists:Track 2 H hist}, which both show at least two modes in their distribution. 
The primary peak of the KDE for each are centred at \(H_{est} = 0.285\) and \(H_{est} = 0.225\) respectively. Both tracks have a smaller, but significant contributions centred at \(H_{est} \approx 0.5\) corresponding to freely diffusive motion. The distributions suggest that the particles are moving in microenvironments with multiple values of $H$. Alternatively, the presence of a physical environment that hinders the motion of the particles with jumps between neighbouring environments suggests that a CTRW-based interpretation may be suitable.\\

Neither T3 or T4 show obvious evidence of caging. The motion of T3 is essentially diffusive while T4 is generally subdiffusive. This is backed up by the distributions in figures \ref{fig:H hists:Track 3 H hist} and \ref{fig:H hists:Track 4 H hist} which are unimodal. The peaks of the KDE are at \(H_{est} = 0.490\) and \(H_{est} = 0.340\) respectively, confirming the clear geometrical changes observed in the tracks. T3 appears to show motion in a Newtonian fluid environment, possibly bulk water, while T4 shows motion in a viscoelastic setting. T3 and T4 seem well modelled by fBm. As such, the moments of their distribution warrant comparison with the performance of the NN to fBm simulations summarised in figure \ref{fig:H simulations}. A cubic spline with increments of \(H\) of 0.01 was used to interpolate between the simulation data points. The point that produces the minimum difference between the spline fit and the value of each moment of the distributions for T3 and T4 was then taken as the estimate for the Hurst exponent in each case. The mean estimates for each track are \(H_{est}^{(3)}=0.523\pm0.009\) and \(H_{est}^{(4)}=0.37\pm0.02\), where the uncertainties are the 95\(\%\) confidence interval. The individual results can be found in table \ref{tab:H sim estimates} in the appendix.\\

\begin{figure*}
\centering
    \subfloat[\label{fig:A tracks:Track 1 A track}]{\includegraphics[width=0.9\columnwidth]{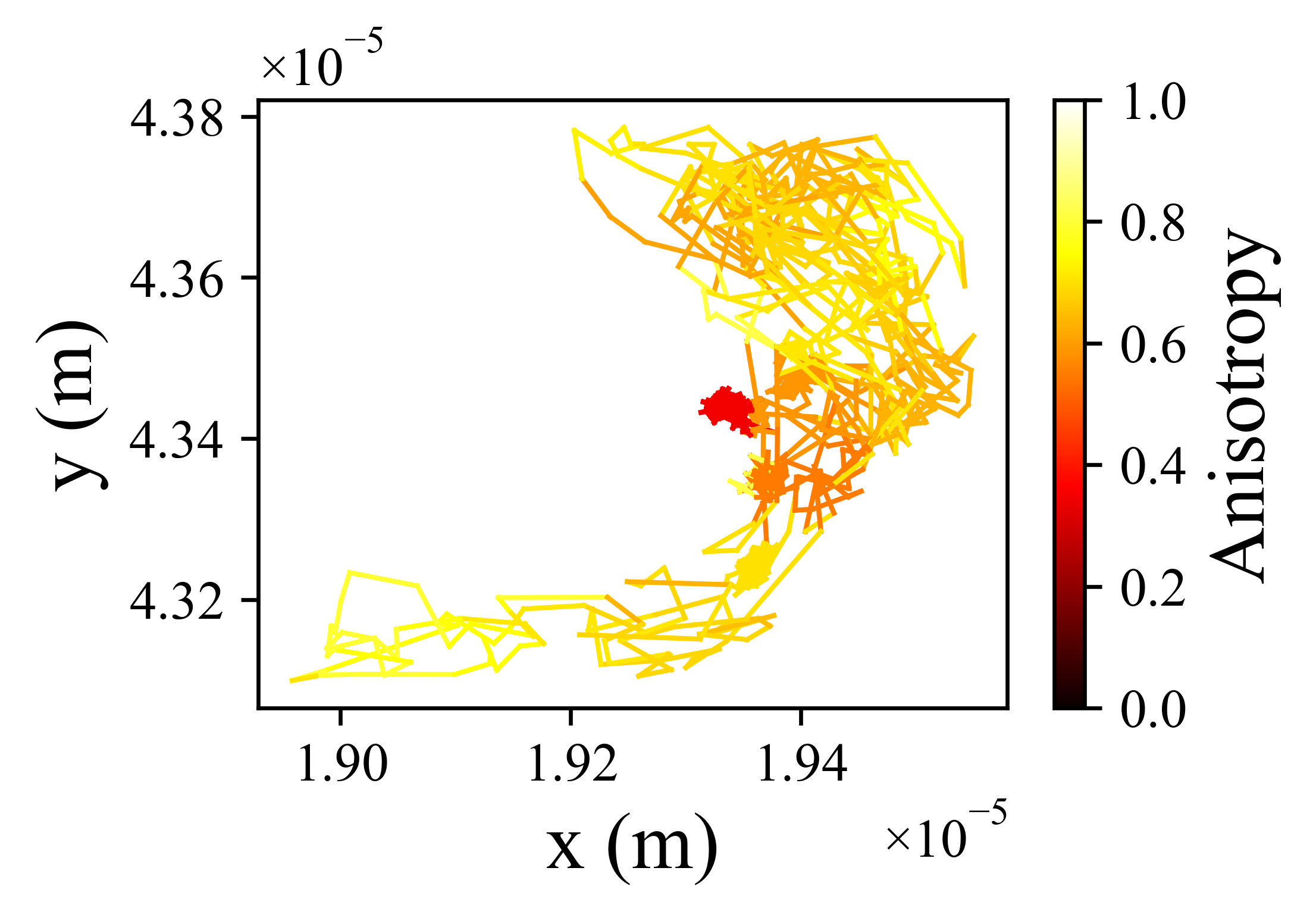}}
    \subfloat[\label{fig:A tracks:Track 2 A track}]{\includegraphics[width=0.9\columnwidth]{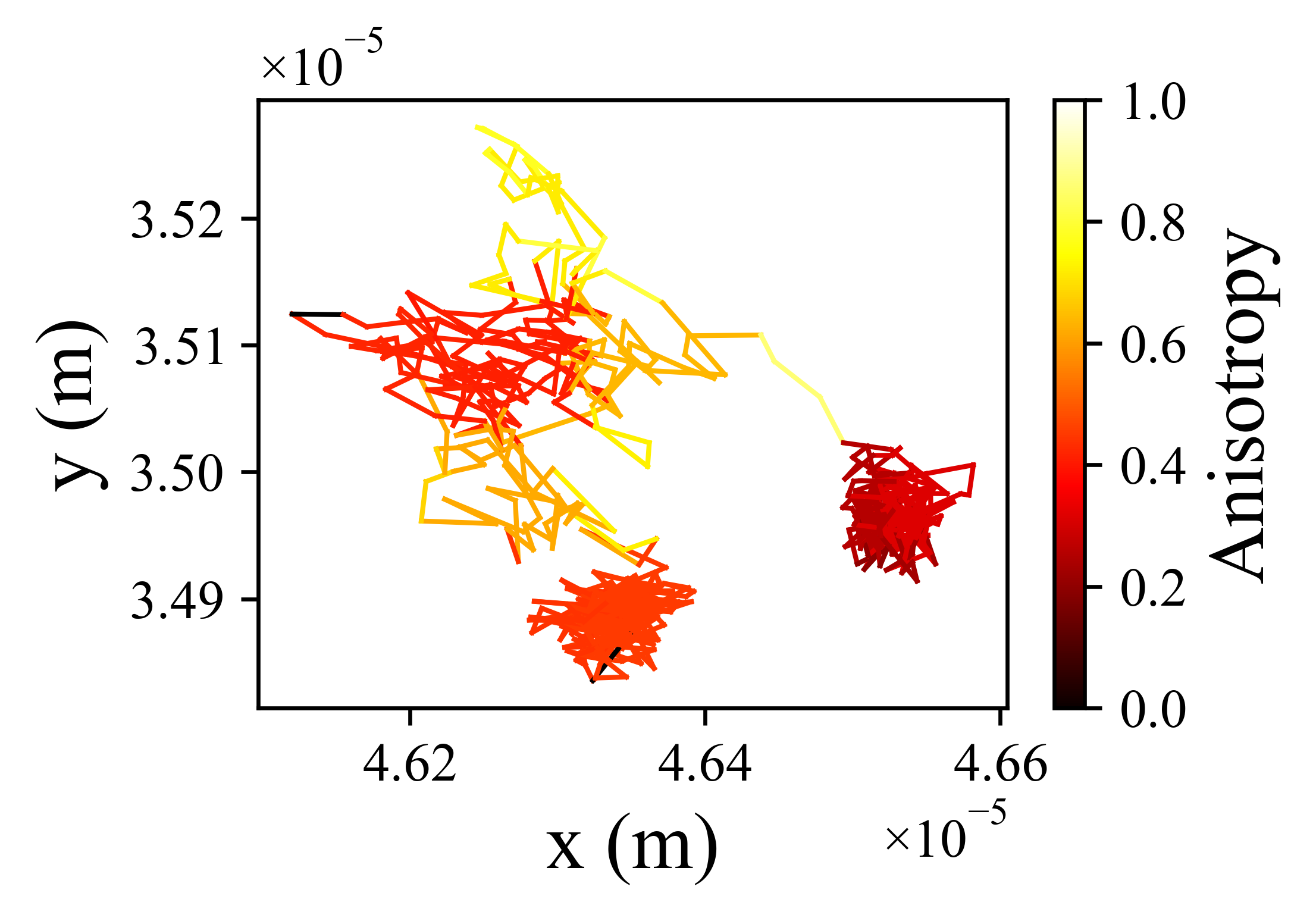}}
    
    \subfloat[\label{fig:A tracks:Track 3 A track}]{\includegraphics[width=0.9\columnwidth]{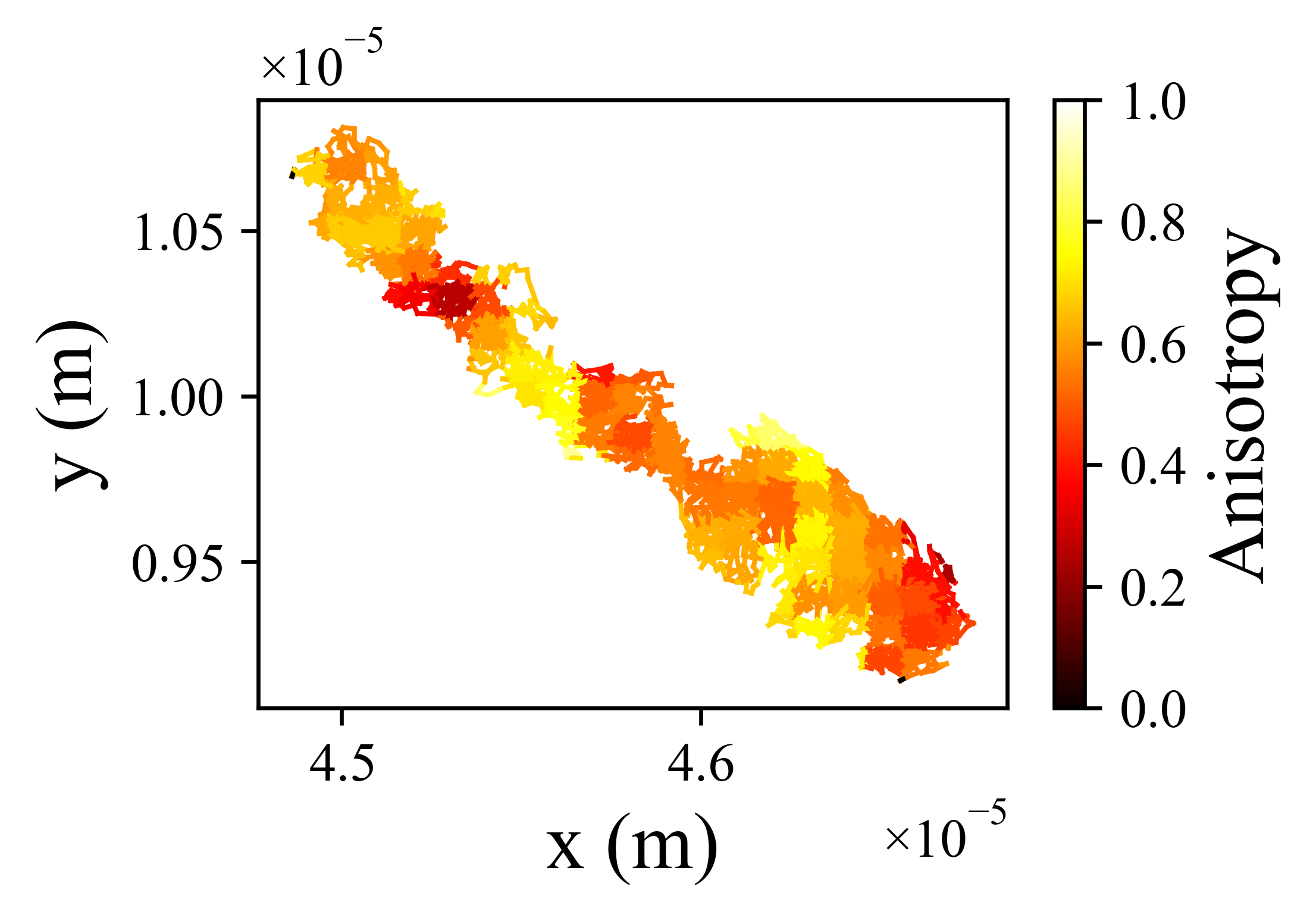}}
    \subfloat[\label{fig:A tracks:Track 4 A track}]{\includegraphics[width=0.9\columnwidth]{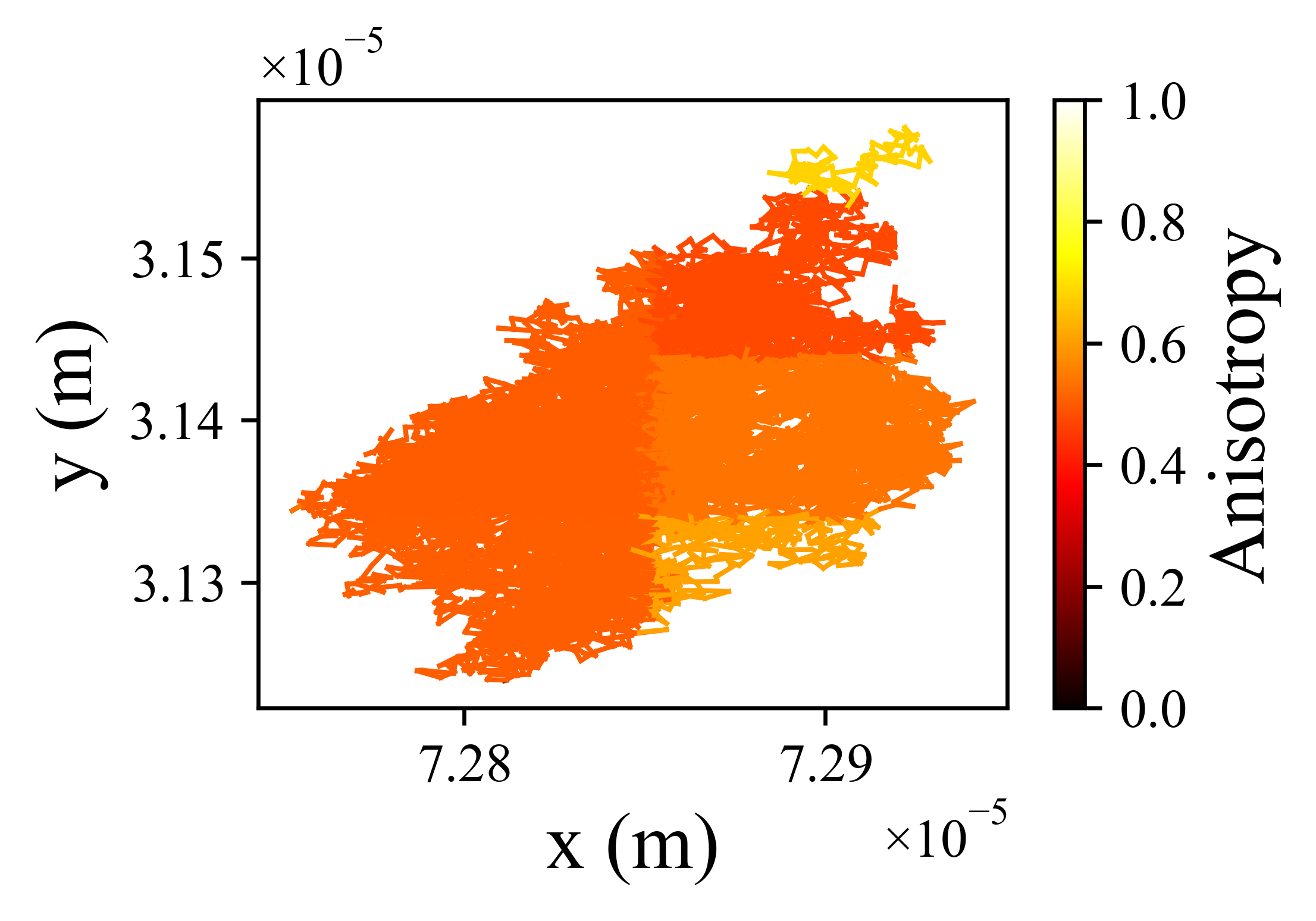}}

    \caption{(a)-(d) show tracks 1-4 respectively with the anisotropy mapped. The anisotropy was calculated via the ratio of the eigenvalues of the radius of gyration tensor, obtained from equation \ref{eq:Anisotropy}. This was for a sliding 45-step window of the track with the value assigned to the point in the centre. The anisotropy was then spatially averaged in the same way as in figure \ref{fig:H tracks} to produce the figures seen here. Anisotropy \(=0\) corresponds to circular symmetry and Anisotropy \(=1\) to motion along a straight line. Plots (a) and (b) show striking similarity to their corresponding plots of H, where the sub-diffusive, caged regions coincide with areas of low anisotropy, and superdiffusive sections coincide with areas of high anisotropy.}
    \label{fig:A tracks}
\end{figure*}

\begin{figure*}
\centering
    \subfloat[\label{fig:A hists:Track 1 A hist}]{\includegraphics[width=0.9\columnwidth]{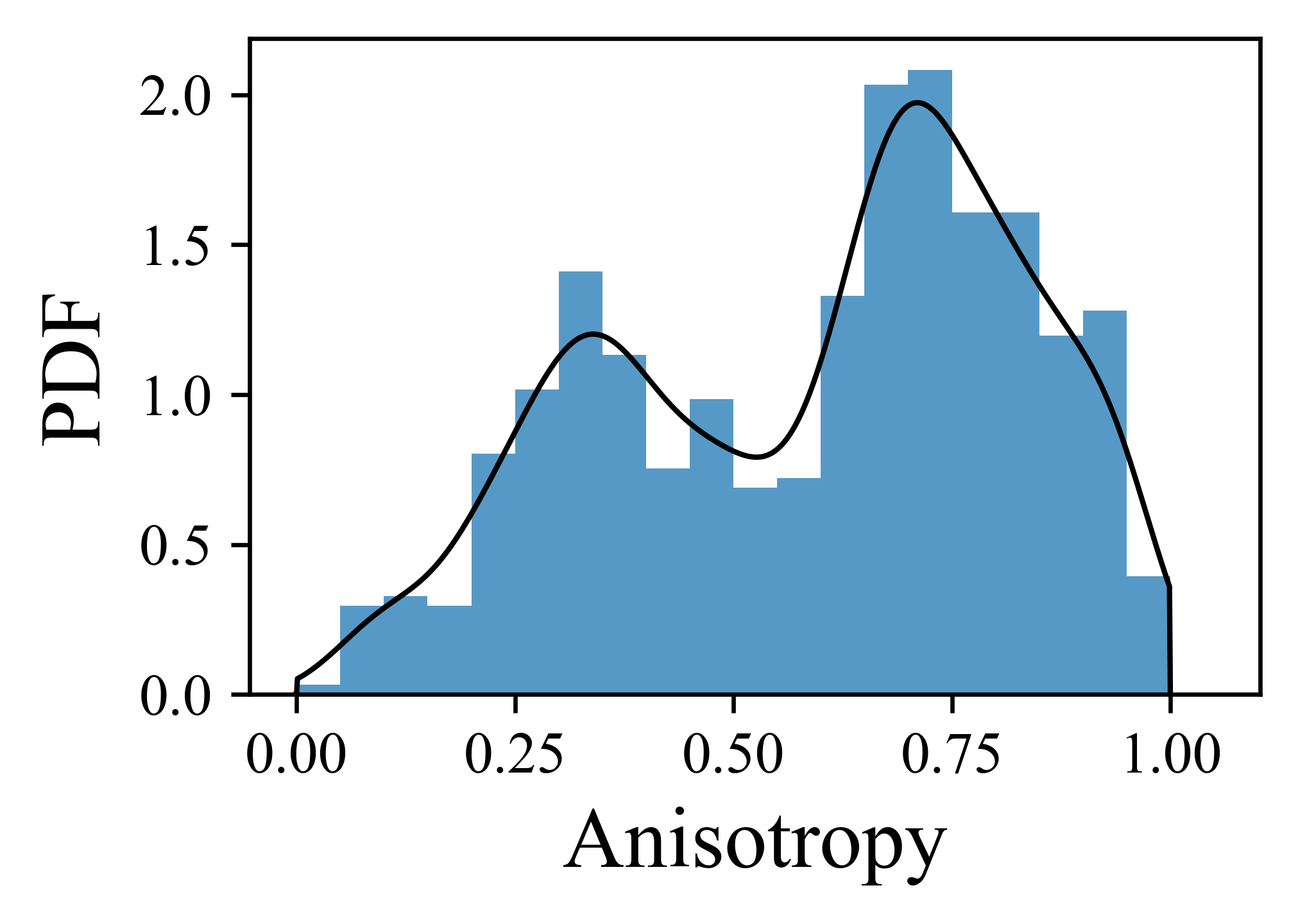}}
    \subfloat[\label{fig:A hists:Track 2 A hist}]{\includegraphics[width=0.9\columnwidth]{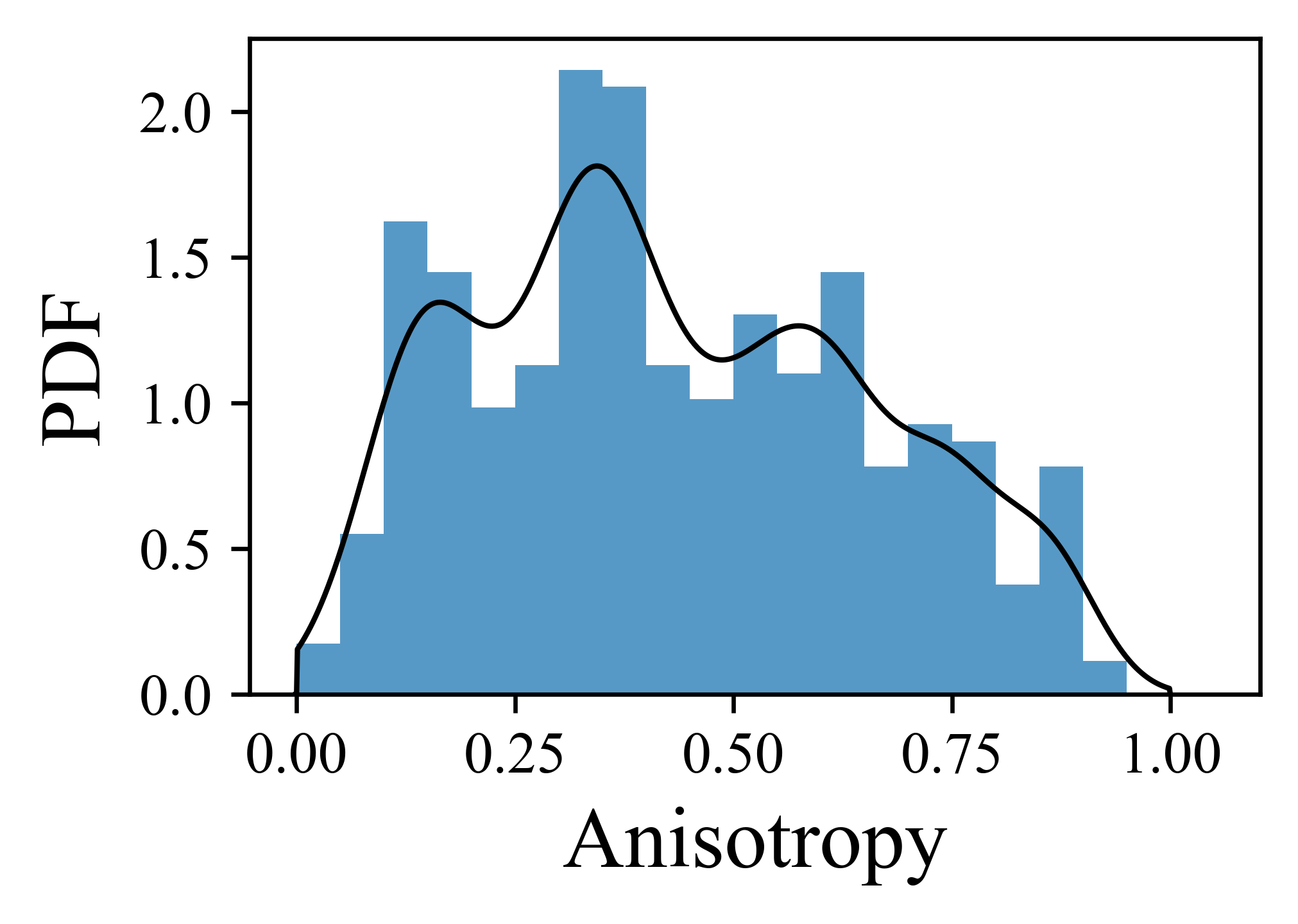}}
    
    \subfloat[\label{fig:A hists:Track 3 A hist}]{\includegraphics[width=0.9\columnwidth]{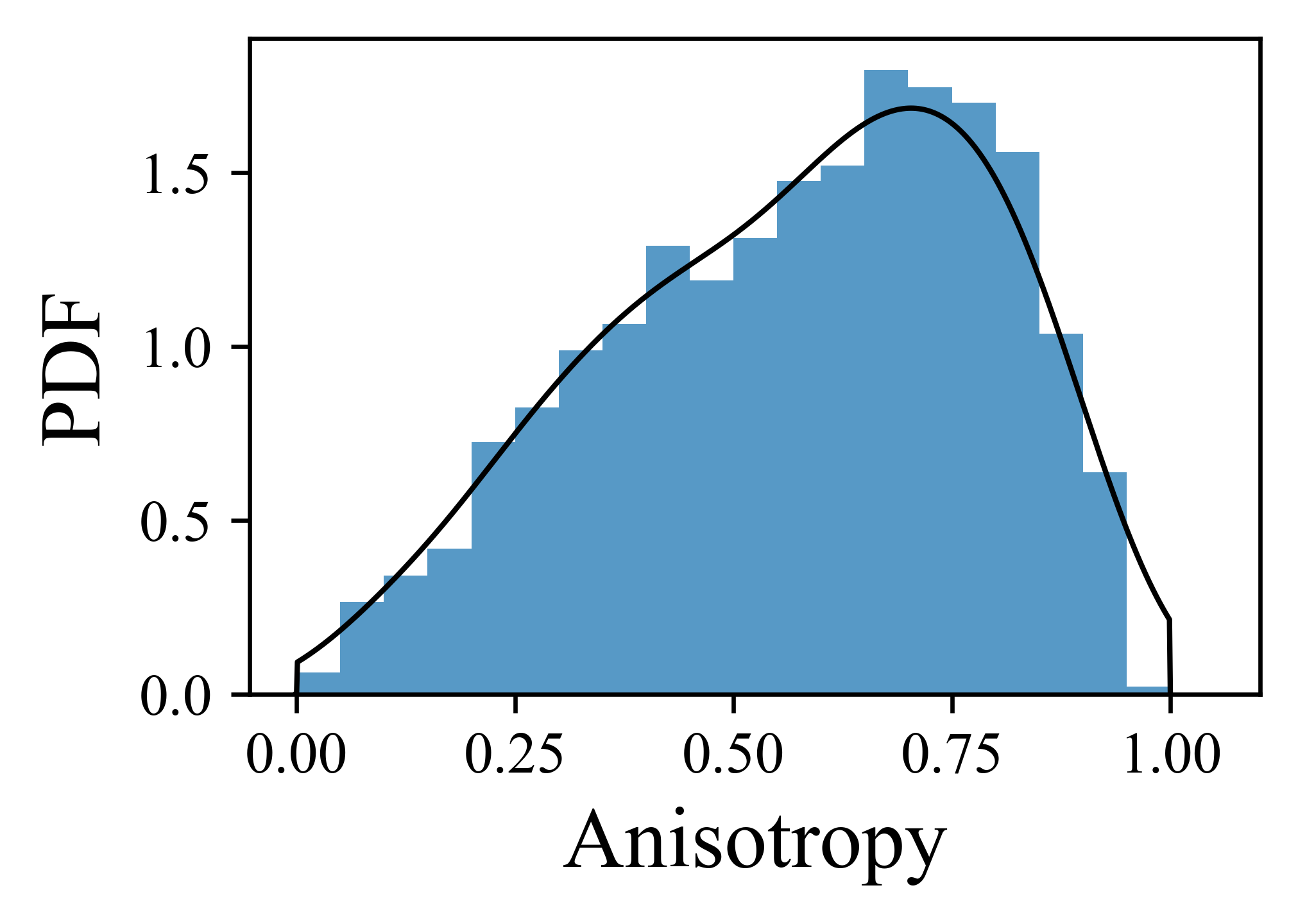}}
    \subfloat[\label{fig:A hists:Track 4 A hist}]{\includegraphics[width=0.9\columnwidth]{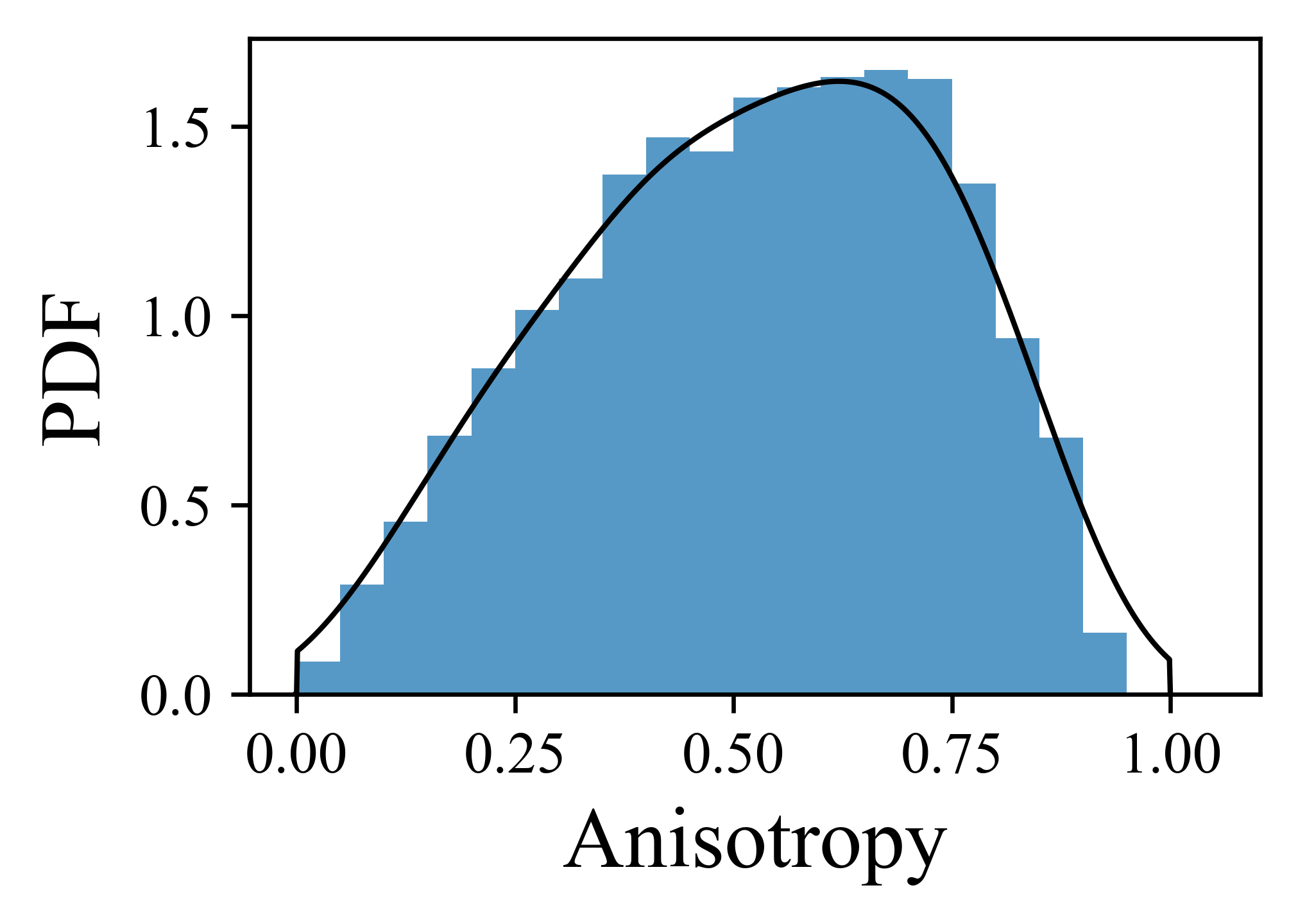}}

    \caption{PDFs of the distribution of anisotropy (equation \ref{eq:Anisotropy}) in tracks 1-4, corresponding to the plots in figure \ref{fig:A tracks}. The orange line shows a Gaussian KDE for each distribution. From simulations, the anisotropy distribution of tracks with a single Hurst exponent are unimodal. The multimodal distributions seen in (a) and (b) therefore highlight the heterogeneity of the working fluid and should be compared with figures \ref{fig:H hists:Track 1 H hist} and \ref{fig:H hists:Track 2 H hist}. In contrast, (c) and (d) both closely resemble the distributions of simulated fBm tracks. This further demonstrates that the majority of the motion in these cases is governed by a single Hurst exponent (see their counterpart in figure \ref{fig:H hists}).}
    \label{fig:A hists}
\end{figure*}

In order to test using the anisotropy as an analytical tool for understanding particle diffusion, T1-4 were treated as in the simulations to produce spatially averaged plots of anisotropy (figure \ref{fig:A tracks}) and the corresponding distributions (figure \ref{fig:A hists}). Gaussian KDEs are again fitted to the plots to aid the eye. Though from the simulations it is clear that anisotropy and Hurst exponent are correlated, the true correlation between the spatially averaged values of \(H_{est}\) and the anisotropy is weak at best. This is most likely owing to the large statistical fluctuations present due to the use of short track sections in the estimation of H and calculation of anisotropy.
In the cages, we see an exception to this where the uniformly subdiffusive value of \(H_{est}\) is accompanied by regions of low anisotropy. On the distributional level, the similarities are much more obvious. The multimodal distributions of \(H_{est}\) for T1 and T2 are replicated in the distributions of anisotropy in figures \ref{fig:A hists:Track 1 A hist} and \ref{fig:A hists:Track 2 A hist}. The anisotropy distributions of T3 and T4 again mimic their \(H_{est}\) counterparts in their unimodal nature and have shapes typical of the distributions obtained in the fBm simulations. The comparison of the four moments of T3 and T4 to simulation can be seen in figure \ref{fig:A simulations}. The same cubic spline method was used to estimate values of \(H\), giving mean values of \(H_{est}^{(3)}=0.467\pm0.009\) and \(H_{est}^{(4)}=0.403\pm0.009\). The full results are summarised in table \ref{tab:Anisotropy H estimates}. These are relatively close to the estimates from the RNN distributions and, to a first approximation, could be used to determine whether a track moving with fBm is subdiffusive. The differences to the RNN most likely stems from the comparatively broad anisotropy distributions and from the imperfect uniformity of the Hurst exponent inherent in real experiments, and in this fluid in particular. In our fluid, there may also be multiple causes of anisotropy, such as physical blocking, adding further deviation from the results of simulations of free motion. 

\subsection{Mean squared displacement}
Non-ergodicity naturally leads to a large spread of \(\overline{r^2}(\Delta)\) about the ensemble mean \cite{Burov2011,He2008}. However, individual TA MSDs can differ significantly from the ensemble in various ways whether the system is ergodic or not. Confinement in the interval \([-L,L]\) results in a plateau of the 1D MSD of fBm tracks at \(\overline{r^2}(\Delta)=\frac{L^2}{3}\) \cite{Jeon2010,Jeon2010_2,Burov2011}. Even though \(\langle r^2(t)\rangle\) for subdiffusive CTRW ensembles are sublinear, \(\langle\overline{r^2}(\Delta)\rangle\), as well as \(\overline{r^2}(\Delta)\) for some tracks, could be linear and so could be mistaken for Brownian motion \cite{He2008,Lubelski2008,Burov2011,Jeon2010}. In the case of confinement, these linear MSDs will transition to a power law with exponent \(1-\alpha\) at \(\frac{L^2}{3}\) \cite{Burov2011}. These plateaus can persist to the ensemble average if the confinement is characteristic of the fluid, such as in networks of actin \cite{Mason1995,Palmer1999,Apgar2000}. MSDs may also be transiently subdiffusive, tending to normal diffusion at long times \cite{Mason1997b,Saxton2007}. Mason et al. (1997) ascribe this to the relaxation of elasticity within a fluid \cite{Mason1997b}. In an fBm system, this suggests that there is a time scale beyond which steps are not correlated. Saxton (2007) shows that this behaviour can also be produced by a particle in a CTRW in which there is a limit on the maximum depth of the traps (rather than an infinite hierarchy) \cite{Saxton2007}.\\
\begin{figure*}
\centering
    \subfloat[\label{fig:Track and ensemble MSDs:Ensemble time MSD}]{\includegraphics[width=0.9\columnwidth]{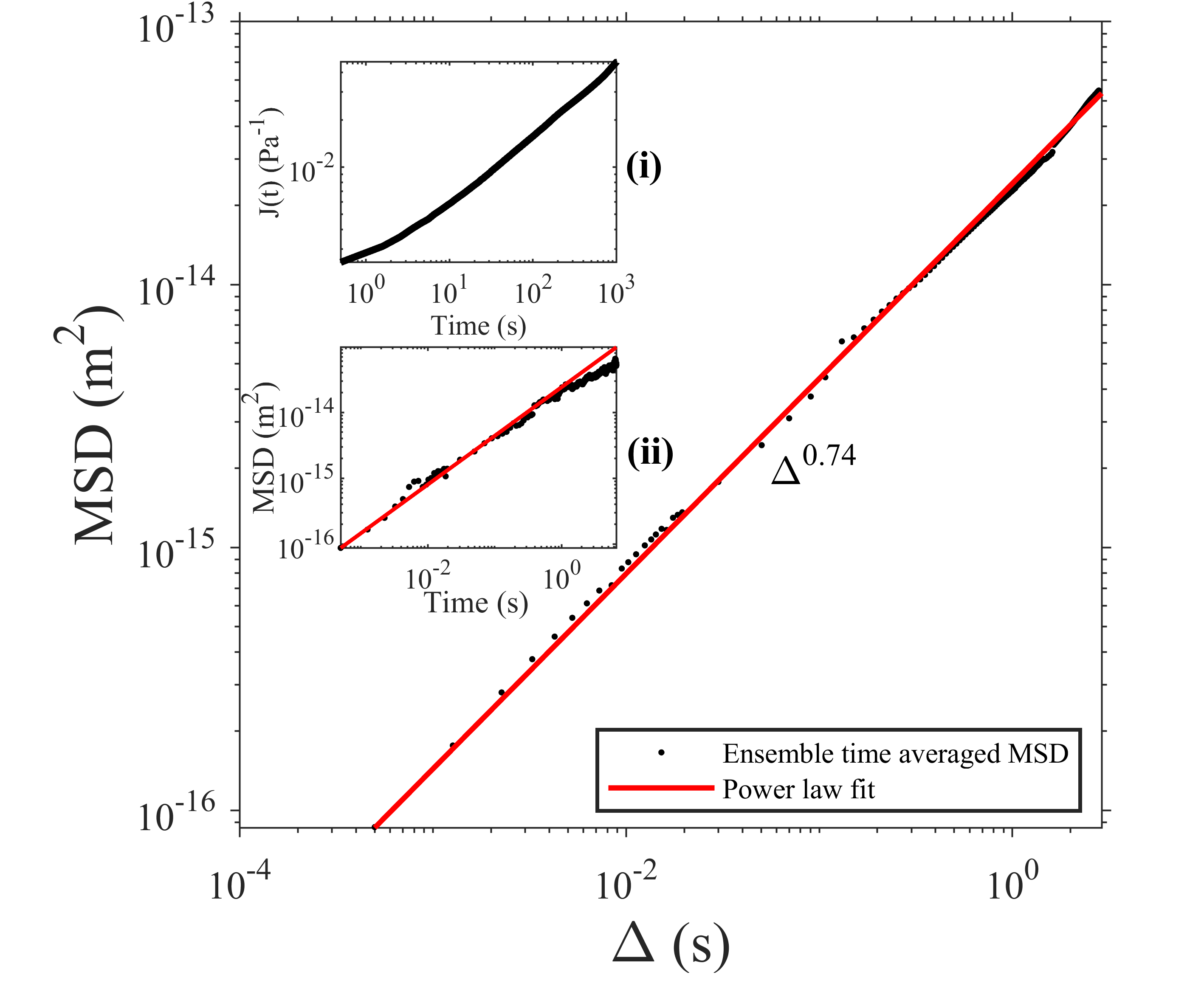}}
    \subfloat[\label{fig:Track and ensemble MSDs:All tracks + PL}]{\includegraphics[width=0.9\columnwidth]{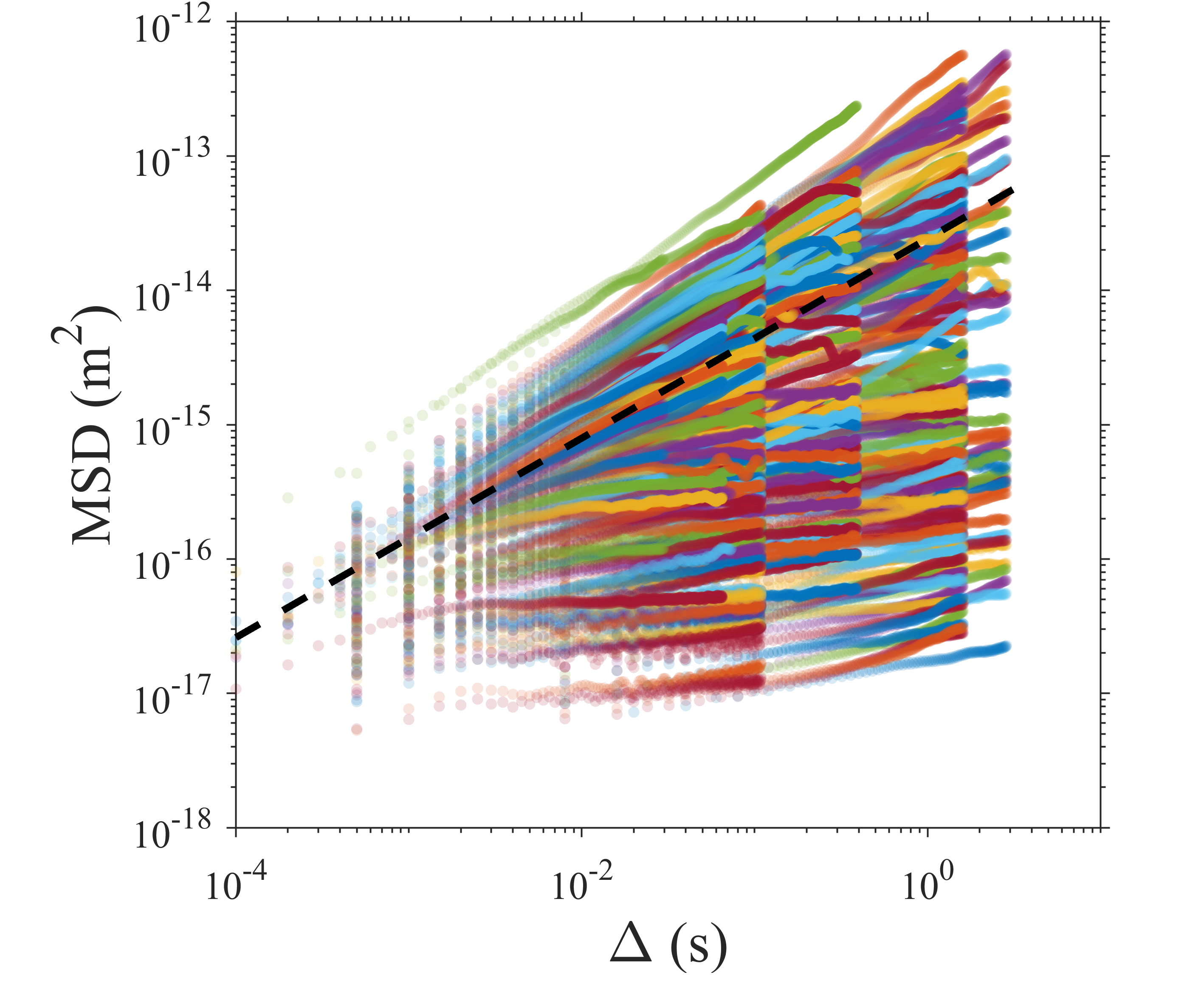}}
    
    \caption{The ensemble average of the time averaged MSDs as a function of time lage, \(\Delta\), is depicted in (a) with a power law fit with an exponent of \(0.74\pm0.02\) overlaid in red. The uncertainty refers to the 95\(\%\) confidence interval. Inset (i) shows the results of a creep-recovery experiment conducted with the same fluid, recreated from Watts Moore et al. 2023 \cite{WattsMoore2023}, also showing power law behaviour. Inset (ii) shows the ensemble averaged MSD with the same power law as the main figure overlaid. The individual time-averaged MSDs as a function of \(\Delta\) across all frame rates are plotted on the same axes in (b), with the power law for the ensemble overlaid as a black dashed line.}
    \label{fig:Track and ensemble MSDs}
\end{figure*}

Our ensemble of particle tracks includes data sets taken at frame rates of 50, 125, 1000, 2000, 5000, and 10000 fps. Even though 414 particle tracks have been used in total, the ensemble MSD proved to be noisy, so the ensemble of the TA MSDs was calculated instead. Track MSDs were binned into time intervals and then averaged, resulting in the plot seen in figure \ref{fig:Track and ensemble MSDs:Ensemble time MSD}. The ensembled data was found to follow a single power law for the whole range of measured time lags: \(\langle \overline{r^2}(\Delta)\rangle=(2.42\pm0.03)\times10^{-14}\Delta^{0.74\pm0.02}\). The uncertainties here correspond to the 95$\%$ confidence interval. The ensemble MSD can be found in inset (ii) to figure \ref{fig:Track and ensemble MSDs:Ensemble time MSD} for reference. A power law MSD corresponds to a power law \(\Tilde{G}(s)\) with the same exponent.

Power laws are common in the rheology of complex fluids and are indicative of a broad, continuous, distribution of time scales responsible for stress relaxation or diffusion \cite{Sollich1997RheologyMaterials,Fielding2000,Bonfanti2020}. They appear in soft glassy rheology \cite{Fielding2000,Sollich1997RheologyMaterials}, which models the disorder present in a fluid as a distribution of energy barriers corresponding to the energy required to reorganise the microstructure. In response to a macroscopic deformation, the barriers are overcome with a distribution of waiting times corresponding to the distribution of energy barriers, thus sharing conceptual similarities with the CTRW.
LGNs such as the one studied in this paper have been shown to exhibit signs of soft glassy rheology \cite{Datta2020CharacterizingNetworks,WattsMoore2023}. 
Inset (i) to figure \ref{fig:Track and ensemble MSDs:Ensemble time MSD} shows the creep compliance, \(J(t)\), for the same fluid sheared in a mechanical rheometer at 3 Pa, below the yield point, recreated from \cite{WattsMoore2023}. The fluid shows power law dynamics but the bulk and micro behaviour don't line up. 
The bulk power law exponent is \(0.437\pm0.003\) \cite{WattsMoore2023} and, if the MSD is converted to \(J(t)\) via equation \ref{eq:compliance}, the two curves are still 3-4 orders of magnitude apart. These differences are stark and most likely stem from the nature of the motion of the tracer particles in the fluid. In the derivation of equations \ref{eq:shear mod} and \ref{eq:compliance}, several assumptions are made. Namely, that fluid forms a continuum, without slip, around the tracer particles, that the tracer particles are larger than the largest elasticity causing structure, and that the motion is ergodic \cite{Mason1995,Mason1997b,Mason2000}. Images taken via polarised light microscopy reveal that there are spherulites present in the fluid with a size of \(\sim 10\; \mu\)m and compound structures that extend over larger length scales still. This is clearly larger than the 0.5 \(\mu\)m diameter of the beads. This fluid is known to exhibit macroscopic wall slip \cite{WattsMoore2023}, so the no-slip condition for the generalised Stokes-Einstein equation may also be violated. The MSDs in figure \ref{fig:Track and ensemble MSDs} are therefore not representative of the bulk dynamics of the fluid and are strongly dependent on local behaviour and heterogeneity on length scales \(\sim0.5 \; \mu\)m.

Figure \ref{fig:Track and ensemble MSDs:All tracks + PL} shows the individual TA MSDs of all of the particle tracks with the ensemble power law overlaid. It is clear that the spread is large, stretching over several orders of magnitude for a given \(\Delta\). Aside from non-ergodic dynamics, the spread of the individual TA MSDs can stem from using short tracks and spatially heterogeneous fluids \cite{Jeon2010}.
The length of the tracks in our ensemble are varied, but most have thousands of steps, so a spread this large is unlikely to originate in the track length. It is far more likely to stem from the spatial heterogeneity revealed in our analysis of \(H\) and the underlying stochastic dynamics.\\

\begin{figure}[b]
    \centering
    \includegraphics[width=0.9\columnwidth]{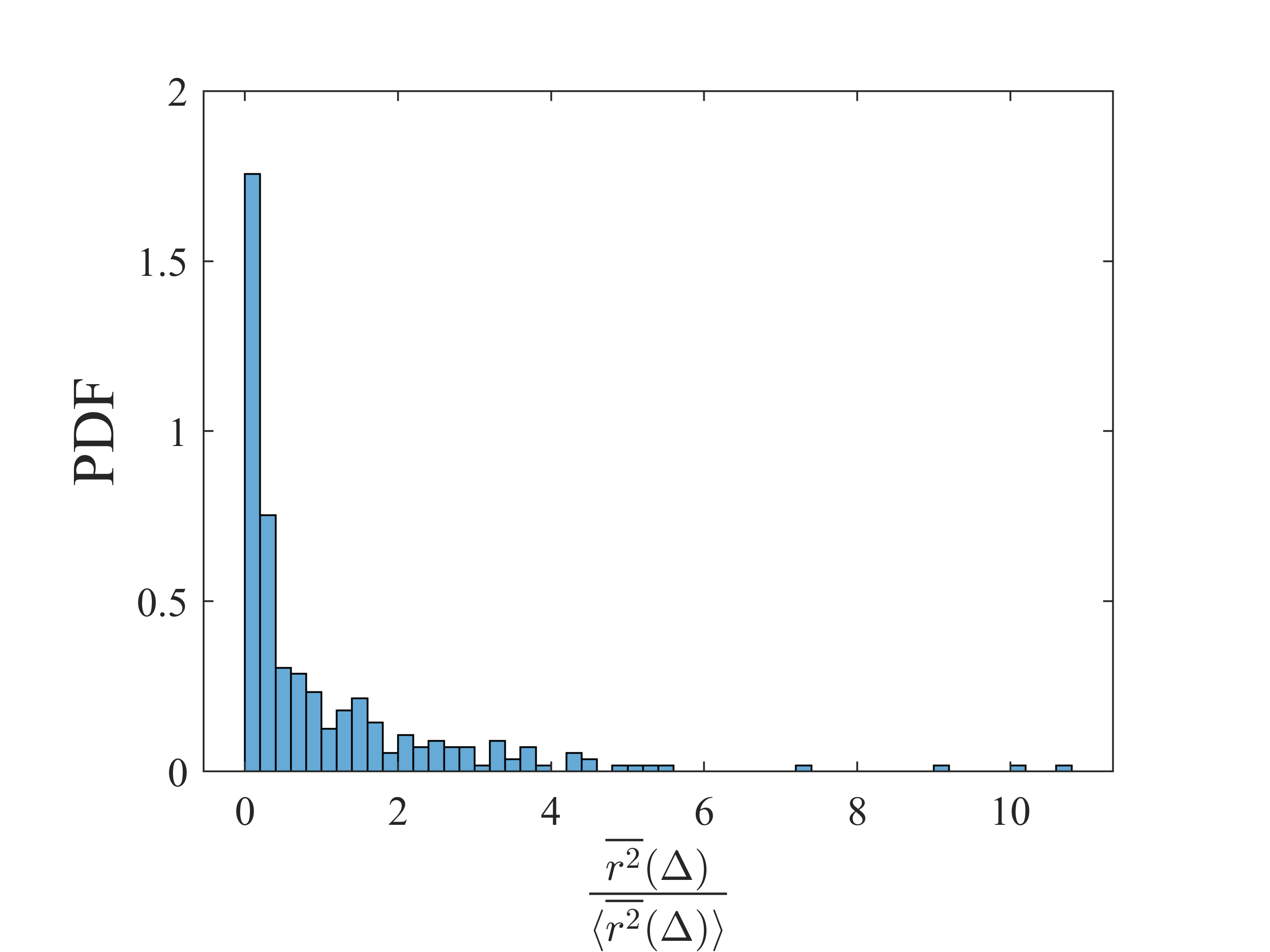}
    \caption{The PDF of the dimensionless quantity, \(\xi=\frac{\overline{r^2}(\Delta)}{\langle\overline{r^2}(\Delta)\rangle}\), that quantifies fluctuations in the MSD at a time lag of \(\Delta=0.02\) s.}
    \label{fig:MSD fluctuations TL=0.02}
\end{figure}

Apgar et al. (2000) and Tseng and Wirtz (2001) used the distribution of the dimensionless quantity, \cite{Apgar2000,Tseng2001}
\begin{equation}
    \xi=\frac{\overline{r^2}(\Delta)}{\langle\overline{r^2}(\Delta)\rangle},
\end{equation} 
to quantify the heterogeneity of probe particle displacements in actin. For CTRWs, the PDF of \(\xi\) should follow a L\'evy distribution \cite{He2008}, while for fBm, the distribution should be Gaussian \cite{Jeon2010}. These relationships hold, even for short tracks, so the distribution can be a key tool in deciphering the dynamics of a system \cite{Jeon2010}. The PDF of \(\xi\) for our tracks at \(\Delta=0.02\) s can be seen in figure \ref{fig:MSD fluctuations TL=0.02}. 
The PDF has a peak at \(\xi=0\) followed by a long-tailed decrease.
The shape of the PDF corresponds to the asymmetry of the spread of individual MSDs around the ensemble in figure \ref{fig:Track and ensemble MSDs}, both of which are hallmarks of non-ergodic dynamics \cite{Jeon2010}. Since the ensemble-time-averaged MSD and the ensemble-averaged MSD have the same power law exponent (see figure \ref{fig:Track and ensemble MSDs}), we conclude that the origin of this non-ergodic behaviour is in the heterogeneity of dynamics \cite{Korabel2023}. However, some non-ergodicity could also stem from the superposition of CTRW-like and fBm-like dynamics.\\
\begin{figure*}
\centering
    \subfloat[\label{fig:Track MSDs:Track 1 MSD}]{\includegraphics[width=0.9\columnwidth]{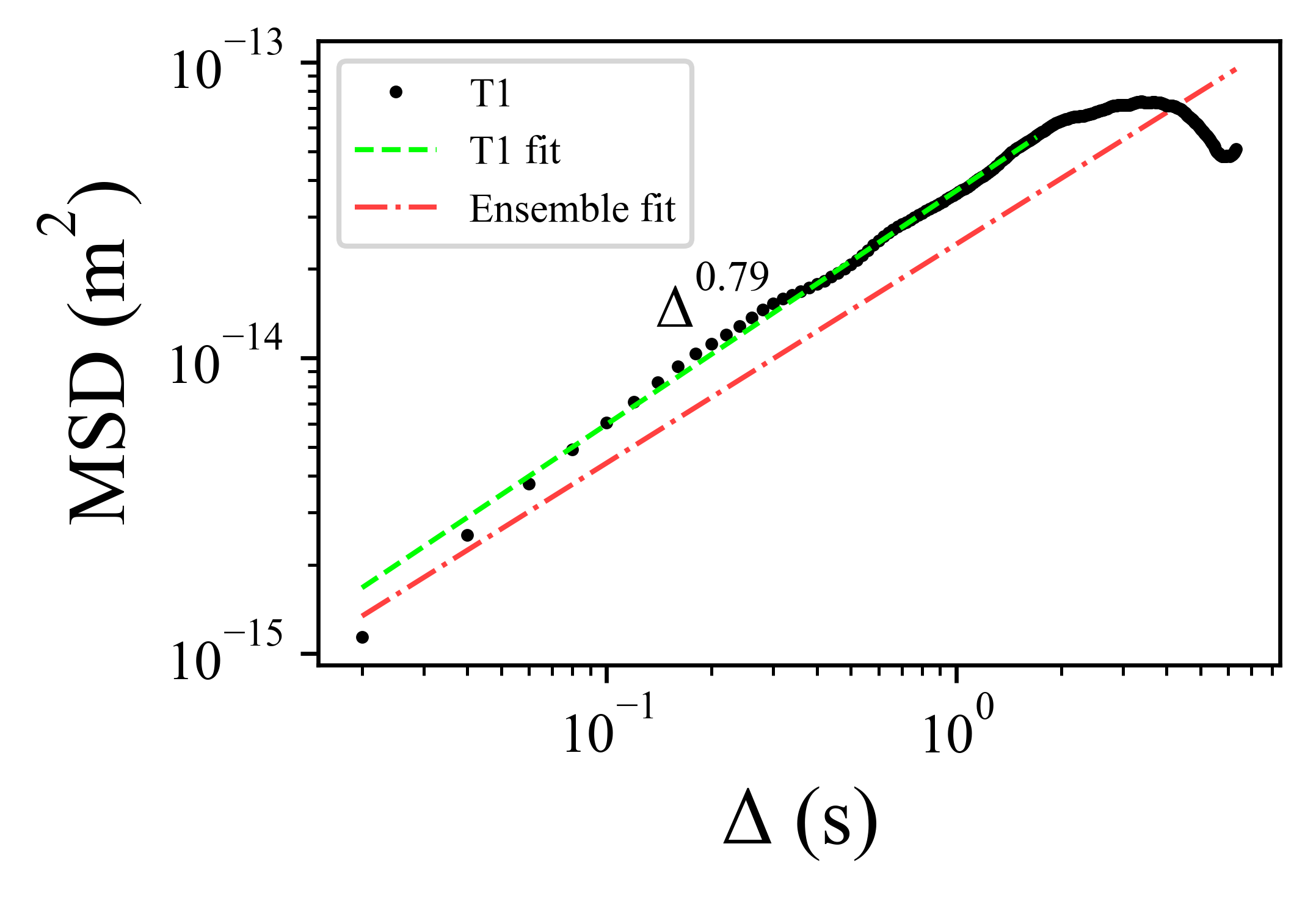}}
    \subfloat[\label{fig:Track MSDs:Track 2 MSD}]{\includegraphics[width=0.9\columnwidth]{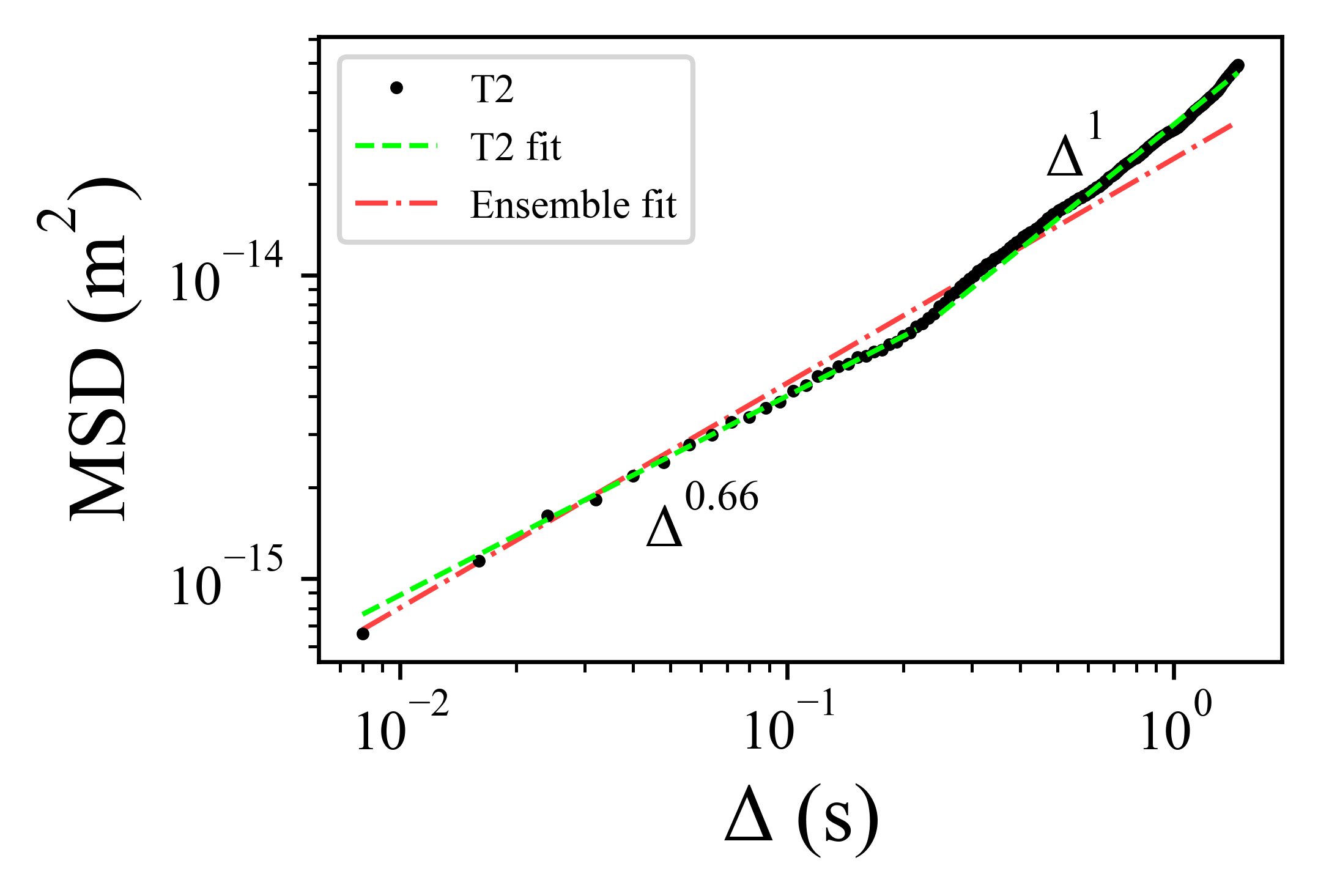}}
    
    \subfloat[\label{fig:Track MSDs:Track 3 MSD}]{\includegraphics[width=0.9\columnwidth]{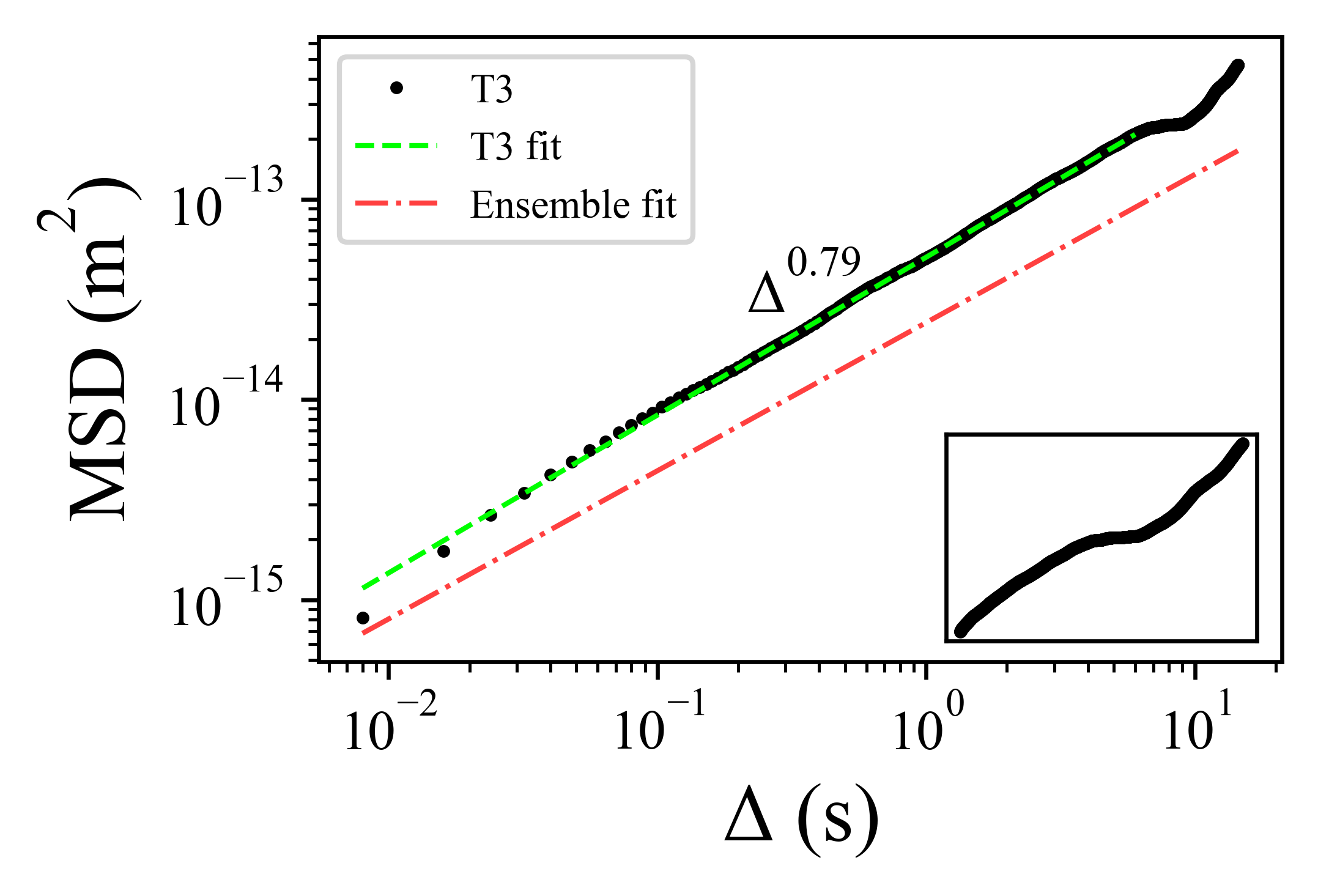}}
    \subfloat[\label{fig:Track MSDs:Track 4 MSD}]{\includegraphics[width=0.9\columnwidth]{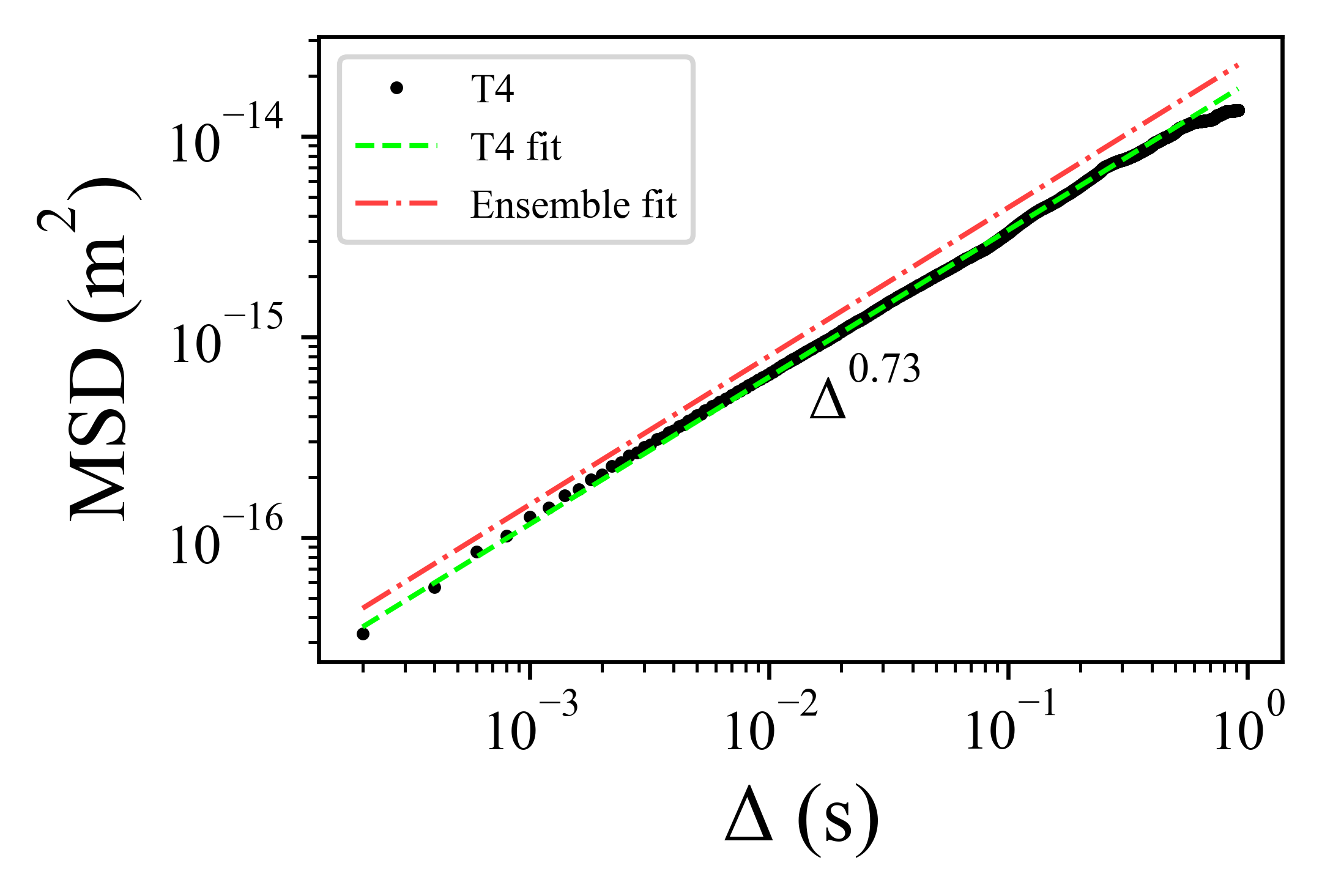}}

    \caption{The TA MSDs of tracks 1-4 as a function of time lag, \(\Delta\), can be seen in figures \ref{fig:Track MSDs:Track 1 MSD}-d. A power has been fitted to at least one section of each curve and is displayed as a red dashed line. A blue dashed line represents the power law of the ensemble average of all the TA MSDs. The inset of \ref{fig:Track MSDs:Track 3 MSD} shows the same MSD plotted on linear scales to highlight the presence of a plateau, often associated with caging phenomena.}
    \label{fig:Track MSDs}
\end{figure*}

Figure \ref{fig:Track MSDs} shows the TA MSDs for T1-4. After the first few points, T1 (figure \ref{fig:Track MSDs:Track 1 MSD}) follows a power law with exponent \(\alpha=0.79\pm0.01\) before plateauing, suggesting the motion is confined. In T2, the particle has three modes: two cages and a free section, where the particle moves diffusively. It spends approximately 2 s in each, which means the majority of its time is spent within the cages. The short \(\Delta\) behaviour is therefore dominated by subdiffusive, intra-cage motion, corresponding to an exponent of \(\alpha=0.66\pm0.02\). At longer \(\Delta\), the average includes jumps from the cages into the free region, meaning the displacement is no longer correlated, producing a linear section in the MSD. If the particle were to continue moving between the same three modes for an extended period of time, the MSD would plateau as the average starts to take into account the distance between the cages.

\begin{figure}[t]
    \centering
    \includegraphics[width=0.9\columnwidth]{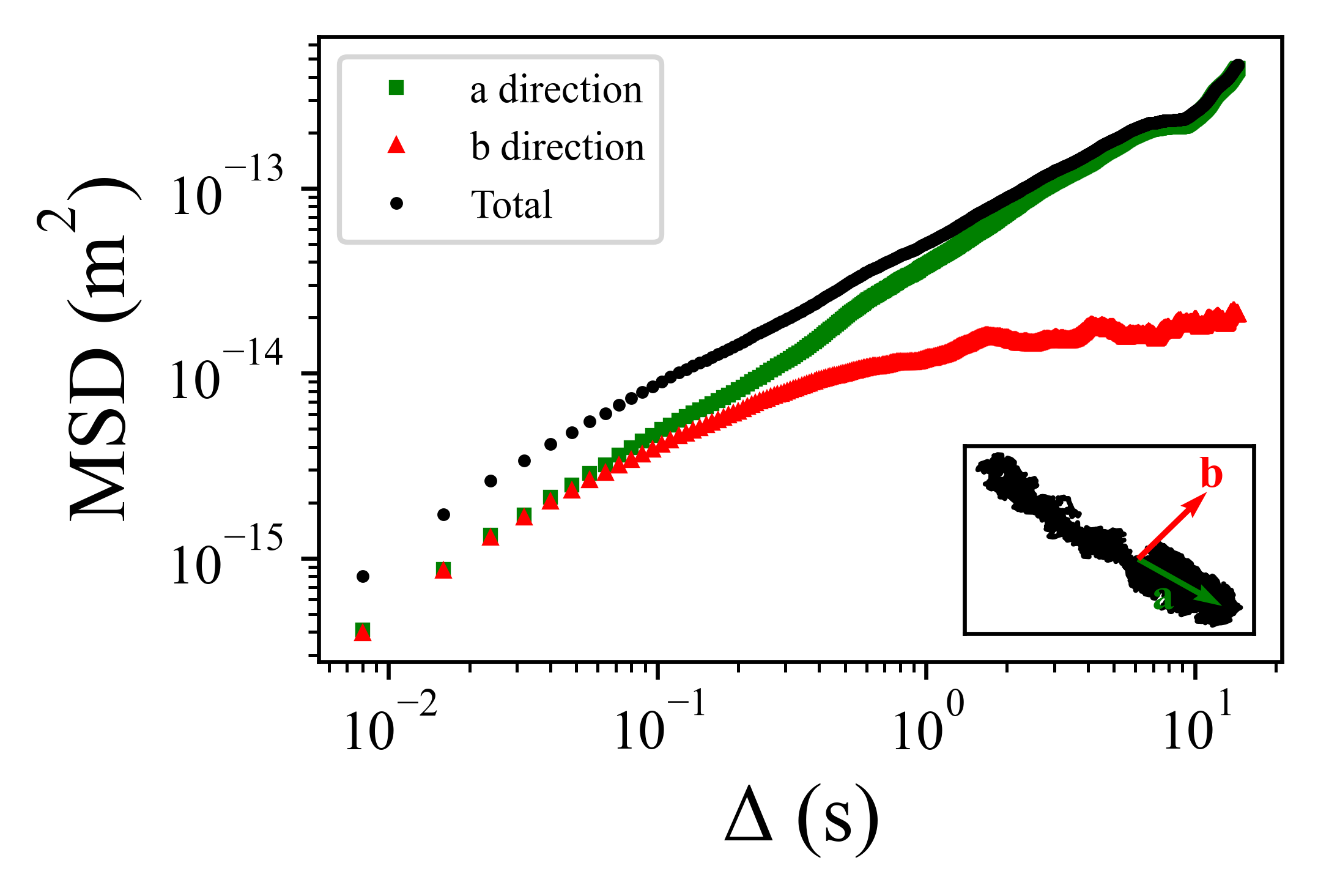}
    \caption{TA MSD as a function of time lag for T3 split into orthogonal components corresponding to the principal axes of the track, a and b. The track with the direction vectors of each axis labelled is shown in the inset.}
    \label{fig:T3 MSD axes}
\end{figure}

The behaviour of T3 again looks subdiffusive, with power law scaling of the form \(\Delta^{0.788\pm0.001}\) for short \(\Delta\), before a plateau is reached. Though this appears to be in conflict with the NN estimates for \(H\) in figure \ref{fig:H hists:Track 3 H hist}, this apparent subdiffusion is actually an expression of confined Brownian motion.
T3 is elongated, which means that we should be able to define its principal axes fairly well using the eigenvectors of equation \ref{eq:GT}. The MSDs in the x and y directions can then be rotated to give the MSDs in the direction of those axes that we will denote a and b through the transformation:
\begin{equation}
    \textbf{EV}\begin{bmatrix}
        x\\y
    \end{bmatrix}=\begin{bmatrix}
        a\\b
    \end{bmatrix},
\end{equation} where \(\textbf{EV}\) is a \(2\times2\) matrix containing the eigenvectors of the gyration tensor. The results of this can be seen in figure \ref{fig:T3 MSD axes}. The MSD for the short axis (MSD\(_b\)) of the track quickly plateaus to approximately \(10^{-14}\) m\(^2\). This corresponds to confinement in the range \(2L\approx0.35\;\mu m\). The MSD for the long axis (MSD\(_a\)) starts off in a similar fashion, but when MSD\(_b\) plateaus, it keeps increasing, tending to the total MSD. This combination of a plateauing and linear MSD combine to create an overall subdiffusive power law. The plateau in the overall MSD at approximately \(2\times10^{-13}\) m\(^2\) corresponds entirely to motion in the a-direction. The length scale of confinement to produce this plateau is \(2L \approx 1.55\;\mu m\).
The values of L obtained from the plateaus correspond well with the dimensions of T3 and the shape of the curves in figure \ref{fig:T3 MSD axes} can easily be reproduced through simulations of Brownian motion with asymmetric physical constraints.
T4 closely follows a power law with \(\alpha=0.733\pm0.003\) for the majority of its MSD. The lack of deviations suggests that the motion is unconstrained and subdiffusive. As with the distribution of \(H\), this suggests that the motion of T3 is well described by fBm. Using the relation \(\alpha=2H\), the exponent suggests \(H\approx0.37\), closely mirroring the estimates based on the distributions of \(H_{est}\) in figure \ref{fig:H hists:Track 4 H hist}.

\subsection{Discussion}
\begin{figure}[b]
    \centering
    \includegraphics[width=0.9\columnwidth]{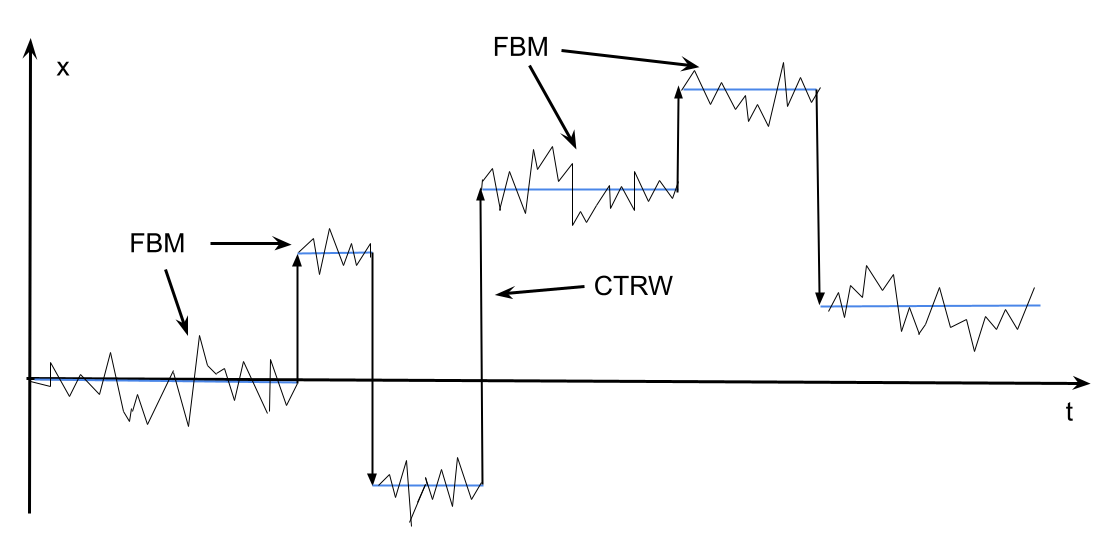}
    \caption{A sketch of the compound dynamics of a particle in one dimension, x, as a function of time, t. The motion is represented by a CTRW process with fBm overlaid. On the microscopic scale only fBm motion can be resolved e.g. with the NN.}
    \label{fig:Sketch}
\end{figure}

There is clearly a large range of dynamics governing diffusion in this LGN that defies categorisation into a single homogeneous stochastic framework. The fluctuations of the time-averaged MSDs about their ensemble averaged quantity point towards heterogeneity \cite{Korabel2023}. Some tracks, including T1 and T2, appear to show elements of a CTRW - long periods on quasi-localised dynamics with transitions between them. At the same time, locally the movement of less than a micron was observed. So, it makes sense to consider a compound process consisting of CTRW and fBm as it is illustrated in \ref{fig:Sketch}.
The confined Brownian motion of T3 and unconfined subdiffusion of T4, well described by fBm, show two more examples of different diffusion dynamics. Given the varied behaviour at play, it is slightly surprising that the ensemble average follows a single, well defined subdiffusive power law. More surprisingly still, the subdiffusion of T4 very closely resembles that of the ensemble, as would be expected from a fluid showing only fBm \cite{Deng2009,Jeon2010}.

The combination of MSD data with RNN estimations for \(H\) has given us greater insight into the stochastic dynamics of individual particle tracks. The multimodal distributions of \(H_{est}\) for T1 and T2 are indicative of an extremely heterogeneous local environment which is responsible for the specific, complex shape of their MSDs. The MSDs of T3 and T4 nominally show similar subdiffusive behaviour for small \(\Delta\). However, their distributions of \(H_{est}\) reveal quite different dynamics: asymmetrically confined Brownian motion for T3, and unconfined fBm for T4. T3 in particular could easily be mistaken for standard fBm subdiffusion if only the MSD were taken into account. The RNN has helped to differentiate between the diverse diffusion processes in our LGN. This is a particularly useful tool for understanding the dynamics in complex, heterogeneous fluids in which dynamics strongly depend on the local environment or in situations where there are few particle tracks available.

\section{Conclusion}
In combination with the analysis of ensemble and time-averaged MSDs, our novel use of an RNN for dynamic segmentation in particle tracking microrheology in a lamellar gel network (LGN) has revealed extreme heterogeneity in the stochastic processes governing its diffusion. Despite the ensemble MSD collapsing to a well-defined subdiffusive power law, CTRW-like behaviour, Brownian motion, and subdiffusion well described by fBm are all present, sometimes coexisiting within the same particle track within \(\sim1\) micron of each other. The non-ergodicity of the ensemble of particles is pronounced, stemming from both the spatial heterogeneity of the LGN microstructure, causing confined motion, and the presence of intrinsically non-ergodic CTRW dynamics. The RNN provides complementary information compared with the MSD, quickly revealing dynamics that may be hard to decipher from single-particle dynamics alone. It has proven to be a useful tool and should become a standard in the microrheologist's toolkit going forward. A possible extension would be to adapt the NN to find the generalized diffusion coefficient which would be expected to be a spatially varying quantity in complex, heterogeneous fluids, such as the LGN in this study.

\begin{acknowledgments}
Many thanks to Daniel Han, Adam Kowalski, and Philip Martin for invaluable work and advice allowing this paper's completion.
This work was made possible with the support of the EPSRC and Unilever under the grant code: EP/R00482X/1.
\end{acknowledgments}

\appendix
\section{Neural network}
\setcounter{table}{0}
\renewcommand{\thetable}{\Alph{section}.\arabic{table}}
\subsection{NN model architecture}\label{sec:NN architecture}
Tables \ref{tab:1 neuron model params}-\ref{tab:Other hyperparams} show a summary of the model architectures of the 1 output neuron and 21 output neuron models that gave the minimum MAE from a Bayesian optimisation routine.
\begin{table}[h]
    \caption{\label{tab:1 neuron model params}%
    Model architecture for RNN with 1 output neuron obtained using Bayesian optimisation.
    }
    \begin{ruledtabular}
    \begin{tabular}{ccccc}
    \textrm{Layer}&
    \textrm{Structure}&
    \textrm{Activation function}&
    \textrm{Neurons}&
    \textrm{Dropout rate}\\
    \colrule
    1 & GRU & selu & 101 & 0.147 \\
    2 & GRU & relu & 94 & 0.233 \\
    3 & BiGRU & softsign & 178 & 0.389\\
    4 & BiLSTM & selu & 261 & 0.184\\
    5 & GRU & selu & 104 & 0\\
    \end{tabular}
    \end{ruledtabular}
\end{table}
The two most common layer types for RNNs are long short-term memory (LSTM) and gated recurrent unit (GRU), which both allow long-term dependencies to be taken into account without the gradient of the loss function diverging during training \cite{Yu2019,Shewalkar2019,Graves2005}. These can be made bidirectional (e.g. BiGRU), meaning that dependencies of the current data point on both past, as in normal RNNs, and future data points in the time series are taken into account \cite{Yu2019,Graves2005}. The activation function of a layer is generally a nonlinear, monotonic, function in which the combined outputs of the neurons from the previous layer are input to activate neurons in the current layer \cite{Tsoi1997}.
The number of neurons is the number of trainable units in a layer. The dropout rate is a mechanism to avoid overfitting and requires a randomly selected proportion of neurons in the layer to ignore during a training batch to prevent the reliance on specific neurons. There are also model-wide parameters that need to be optimised, namely the model learning rate, and batch size. The former is related to the size of the steps taken by the model when minimising the loss 
\begin{table}
    \caption{\label{tab:multi neuron model params}
    Model architecture for RNN with 21 output neurons obtained using Bayesian optimisation.}
    \begin{ruledtabular}
    \begin{tabular}{ccccc}
    \textrm{Layer}&
    \textrm{Structure}&
    \textrm{Activation function}&
    \textrm{Neurons}&
    \textrm{Dropout rate}\\
    \colrule
    1 & BiLSTM & selu & 172 & 0.577 \\
    2 & BiLSTM & relu & 476 & 0.229 \\
    3 & BiLSTM & tanh & 225 & 0.359\\
    4 & GRU & tanh & 165 & 0.160\\
    5 & BiLSTM & softplus & 189 & 0\\
    \end{tabular}
    \end{ruledtabular}
\end{table}
\begin{table}[t]
    \caption{\label{tab:Other hyperparams}
    Number of output neurons, output activation function (AF), learning rate (LR), and batch size (BS) for each RNN model.}
    \begin{ruledtabular}
    \begin{tabular}{cccc}
    \textrm{Output neurons}&
    \textrm{Output AF}&
    \textrm{LR}&
    \textrm{BS}\\
    \colrule
    1 & linear & 0.001 & 37 \\
    21 & softmax & 0.0008 & 44\\
    \end{tabular}
    \end{ruledtabular}
\end{table}function and the latter is the number of data sets from the training set used in a single iteration.

In order to ensure that each model can be accurately compared, they must be trained with the same tracks and on the same PC. The PC used was a Microsoft Azure virtual machine with a Linux operating system (Ubuntu 20.1), 4 vcpus, and 32 GB memory. When used in combination with the 1- neuron model, small changes were manually made to the biases on the output neurons of the 21-neuron model to make sure that no value of \(H\) were over or under predicted.

\subsection{Hurst exponent estimates}
Tables \ref{tab:H sim estimates} and \ref{tab:Anisotropy H estimates} summarise the estimations of the Hurst exponent, \(H\), for track 3 and 4 from the moments of the distributions of the anisotropy and \(H\) simulations. A cubic spline interpolation was used to find the corresponding value of \(H_{sim}\) that would produce a distribution with the same moments. The results of this can be seen in figures \ref{fig:H simulations} and \ref{fig:A simulations}.
\begin{table}[h]
    \caption{\label{tab:H sim estimates}
    Estimates for \(H\) based on the mean, skewness, and KDE peak (\(H_m,\; H_s,\) and \(H_{kde}\)) of the distributions of \(H_{est}\) for T3 and T4.}
    \begin{ruledtabular}
    \begin{tabular}{cccc}
    \textrm{Track}&
    \textrm{\(H_{m}\)}&
    \textrm{\(H_{s}\)}&
    \textrm{\(H_{kde}\)}\\
    \colrule
    3 & 0.53 & 0.52 & 0.52 \\
    4 & 0.35 & 0.38 & 0.37\\
    \end{tabular}
    \end{ruledtabular}
\end{table}
\begin{table}[h]
    \caption{\label{tab:Anisotropy H estimates}
    Estimates for the Hurst exponent based on the mean, skew and KDE peak (\(H_{A,m},\; H_{A,s}\), and \(H_{A,kde}\)) of the anisotropy distributions of T3 and T4.}
    \begin{ruledtabular}
    \begin{tabular}{cccc}
    \textrm{Track}&
    \textrm{\(H_{A,m}\)}&
    \textrm{\(H_{A,s}\)}&
    \textrm{\(H_{A,kde}\)}\\
    \colrule
    3 & 0.47 & 0.46 & 0.47 \\
    4 & 0.40 & 0.40 & 0.41 \\
    \end{tabular}
    \end{ruledtabular}
\end{table}
\clearpage

\bibliography{references}

\end{document}